\documentclass[article,aps,prl,twocolumn,superscriptaddress]{revtex4-2} 
\usepackage{amsmath,amssymb,graphicx,bm,gensymb,url}
\usepackage[most]{tcolorbox}
\usepackage{subcaption}
\usepackage[colorlinks=true,linkcolor=blue,citecolor=blue,urlcolor=blue]{hyperref}
\begin{document}

\title{\textit{C-BerryTrans}: A C++ code for \textit{first-principles} calculation of Berry-curvature-driven anomalous Hall and Nernst conductivities.}

\author{Vivek Pandey}
\altaffiliation{vivek6422763@gmail.com}
\affiliation{School of Physical Sciences, Indian Institute of Technology Mandi, Kamand - 175075, India}
\author{Sudhir K. Pandey}
\altaffiliation{sudhir@iitmandi.ac.in}
\affiliation{School of Mechanical and Materials Engineering, Indian Institute of Technology Mandi, Kamand - 175075, India}

\date{\today}

\begin{abstract}
We present \textit{C-BerryTrans}, a C++ code designed for \textit{first-principles} calculations of Berry-curvature-driven transverse transport properties, namely the anomalous Hall conductivity (AHC) \textit{i.e.}, $\sigma_{\mu \nu}^{AHC}$ and anomalous Nernst conductivity (ANC) \textit{i.e.}, $\alpha_{\mu \nu}^{ANC}$. The code directly extracts eigenvalues and momentum-matrix elements from WIEN2k calculations and evaluates the Berry curvature ($\boldsymbol\Omega$) using a Kubo-like formalism, thereby avoiding the interpolation errors inherent in Wannier-based approaches. To ensure computational efficiency, \textit{C-BerryTrans} parallelizes $\boldsymbol\Omega$ evaluation over \textbf{\textit{k}}-points using OpenMP and stores band-resolved curvature data in binary format, significantly reducing memory usage. This design enables rapid post-processing of AHC and ANC over a wide range of temperature ($T$) and chemical potential ($\omega$) values in a single run. The code has been benchmarked on well-studied ferromagnetic materials- Fe, Fe$_3$Ge, Pd, Fe$_3$Al, and Co$_2$FeAl. For Fe, the $\sigma_{xy}^{AHC}$ is obtained to be $\sim$775 ($\sim$744) $S/cm$ at 0 (300) $K$. In case of Fe$_3$Ge, the calculated value of $\sigma_{xy}^{AHC}$ is found to be 311 $S/cm$ at 300 $K$. Nextly, for Co$_2$FeAl, the magnitude of computed value of $\sigma_{xy}^{AHC}$ at 2 $K$ is found to be $\sim$56 $S/cm$. Moving further, the room temperature magnitude of $\alpha_{xy}^{ANC}$ for Pd is obtained to be $\sim$0.97 $AK^{-1}m^{-1}$. In case of Fe$_3$Al, the maximum magnitude of $\alpha_{xy}^{ANC}$ for $T\leq$500 $K$ is computed as $\sim$2.83 $AK^{-1}m^{-1}$. Lastly, for Co$_2$FeAl, the value of $\alpha_{xy}^{ANC}$ is obtained to be $\sim$0.10 $AK^{-1}m^{-1}$ at 300 $K$. These results show  excellent agreement with previously reported data. With its accuracy, scalability, and user-friendly workflow, \textit{C-BerryTrans} provides a powerful tool for exploring $\boldsymbol\Omega$-driven transport phenomena and is well suited for high-throughput materials discovery. The code further enables the evaluation of $\boldsymbol\Omega$-derived AHC/ANC contributions along user-defined high-symmetry \textbf{\textit{k}}-point paths. This provides valuable microscopic insight into how specific band-structure features contribute to $\boldsymbol\Omega$-driven AHC/ANC. Additionally, the code is equipped with a visualization module that allows analysis of \textbf{\textit{k}}-point contributions to AHC or ANC in any material. This further enhances its capability for exploring topological materials.\\

\textbf{Program summary -}\\
\textit{Program Title}: \textit{C-BerryTrans}\\
\textit{Program Files doi}:\\
\textit{Developer’s repository link}: \url{https://sourceforge.net/projects/c-berrytrans/}\\
\textit{Licensing provisions}: GNU General Public License 3.0\\
\textit{Programming language}: C++\\
\textit{External routines/libraries}: cmath, vector, chrono, omp.h, cstdlib, string\\
\textit{Nature of problem}: The code evaluates the AHC and ANC within the \textit{ab initio} framework. It first computes the $\boldsymbol\Omega$ using the Kubo formula, where the eigenvalues and momentum-matrix elements are directly extracted from \textit{first-principles} calculations carried out with WIEN2k. The obtained $\boldsymbol\Omega$ is then integrated over the entire Brillouin zone (BZ). For AHC, the integration is weighted by the Fermi–Dirac distribution to include the effect of electronic occupation at different values of $\omega$ and $T$. For ANC, the same formalism is extended by incorporating the entropy function weighting, thereby capturing the thermoelectric response arising from the $\boldsymbol\Omega$.

\end{abstract}

\maketitle

\section{Introduction}

$\boldsymbol\Omega$ plays a central role in determining the transverse transport properties of crystalline solids and has significant technological applications\cite{korrapati2024nonlinear,okyay2022second,sodemann2015quantum,kim2021vertical}. Two of the most important $\boldsymbol\Omega$-driven responses are the AHC and the ANC. AHC arising from a non-vanishing $\boldsymbol\Omega$ leads to dissipationless Hall currents, which are very useful for spintronics application\cite{fukami2020antiferromagnetic,miah2007observation} and quantum topological phenomena\cite{bradlyn2017topological,wieder2020magnetic}. Apart from this, such intrinsic AHC has been found to have numerous applications in the diluted magnetic semiconductors\cite{jungwirth2002anomalous}. A recent experimental study revealed that symmetry elements of Heusler magnets can be changed such that the $\boldsymbol\Omega$ and all the associated properties are switched while leaving the magnetization unaffected. This helps in tuning the AHC from 0 to $\sim$1600 $Ohm^{-1}cm^{-1}$ in full Heusler alloys Co$_2$MnGa \& Mn$_2$CoGa\cite{manna2018colossal}. Such tunability of AHC is expected to enhance the practical applications of these compounds. Moving to the case of anomalous Nernst effect (ANE) which gives rise to ANC, the phenomenon is very efficient in thermal energy harvesting. In this direction, the effect has been recently found to produce high power density of around 13$\pm$2 W/cm$^3$ in nanofabricated device\cite{lopez2022high}. In addition to this, the phenomenon is useful for designing heat flux sensors\cite{zhou2020heat}. The study of interplay between the spin Seebeck effect and ANE revealed the importance of ANE in spin-caloritronics\cite{huang2011intrinsic,bosu2012thermal,weiler2012local}. Also, ANE in a material enables its use in novel magnetic imaging microscopy techniques that rely on thermal gradients\cite{weiler2012local,reichlova2019imaging,janda2020magneto,johnson2022identifying}. Furthermore, ANE and AHE are useful for the characterization of topological nature of charge carriers in magnetic WSMs. It is worth noting that unlike AHC, which depends on overall integration of all the occupied bands, ANC arises only from the bands which contribute in the vicinity of the Fermi energy\cite{ikhlas2017large,xiao2006berry,xiao2010berry}. This makes the ANE measurement highly useful to characterize the $\boldsymbol\Omega$ distribution close to Fermi energy and to verify the possibility of Weyl phase. These discussions provide the motivation for efficient and accurate computation of these $\boldsymbol\Omega$-driven transport properties.

Theoretically, AHC is computed by integrating $\boldsymbol\Omega$ over all the bands at each \textbf{\textit{k}}-points across the BZ, with contributions weighted by the Fermi-Dirac distribution to reflect electronic occupation at a given chemical potential ($\omega$) and temperature ($T$). For ANC computation, the work of Xiao \textit{et. al.} suggests that it can be done in a similar fashion of $\boldsymbol\Omega$ integration over the entire BZ but weighted by the entropy function. The expression for AHC ($\sigma_{\mu \nu}^{\mathrm{AHC}}$) is given by\cite{yao2004first,xiao2010berry,ernst2019anomalous,helman2021anomalous,wang2006ab}: 
\begin{multline}
\sigma_{\mu \nu}^{\mathrm{AHC}}=-\frac{e^2}{\hbar\epsilon_{\mu\nu\xi}} \frac{1}{(2 \pi)^3} \sum_n \int_{\mathrm{BZ}} f_n(\textbf{\textit{k}}) \Omega_{\mu \nu}^n(\textbf{\textit{k}}) d \textbf{\textit{k}}
\label{eqAHE}
\end{multline}
while for ANC ($\alpha_{\mu \nu}^{\mathrm{ANC}}$), it is given as\cite{PhysRevB.85.012405,xiao2010berry} 
\begin{multline}
\alpha_{\mu \nu}^{\mathrm{ANC}} = -\frac{1}{T} \frac{e}{\hbar} \sum_n \int \frac{d^3k}{(2\pi)^3} \, \Omega_{\mu \nu}^n(\textbf{\textit{k}}) \left[ (E_n - E_F) f_n(\textbf{\textit{k}}) \right. \\
\left. + \, k_B T \ln \left(1 + \exp\left( \frac{E_n - E_F}{-k_B T} \right) \right) \right]
\label{eqANE}
\end{multline}
Here, $\epsilon_{\mu\nu\xi}$ is the Levi-Civita symbol, $f_n(\textbf{\textit{k}})$ is the Fermi-Dirac distribution function and $\Omega_{\mu \nu}^n(\textbf{\textit{k}})$ is the component of $\boldsymbol\Omega$ for band index $n$ and at crystal momentum \textbf{\textit{k}}. Furthermore, \textit{e}, $\hbar$, $k_B$ and $T$ denote the electronic charge, reduced Planck\textquoteright s constant, Boltzmann constant and temperature, respectively. The symbols $\mu$, $\nu$ and $\xi$ stand for any of the $x$, $y$ and $z$ cartesian direction. From the well-known Kubo formula for computing $\Omega_{\mu \nu}^n(\textbf{\textit{k}})$, it is known that the term is highly sensitive to the details of electronic structure of the material under study, especially on the energy eigenvalues and momentum-matrix elements associated with the particular (\textit{n},\textbf{\textit{k}}) Bloch state\cite{chang1996berry,sundaram1999wave,onoda2002topological,jungwirth2002anomalous}. Thus, the accuracy of the calculated properties depends on how precise and accurate computation of energy eigenvalue and the momentum-matrix elements of (\textit{n},\textbf{\textit{k}}) state is done. This in turn depends on how accurately the (\textit{n},\textbf{\textit{k}}) is determined. Moreover, an accurate and efficient numerical approach is required to evaluate the integral over a dense \textbf{\textit{k}}-point grid. It is important to mention here that according to the classification of Onoda, Sugimoto, and Nagaosa \cite{onoda2008quantum}, this intrinsic part (given by Eqs. \ref{eqAHE} \& \ref{eqANE}) is physically meaningful only in the clean to moderately dirty regimes, where quasiparticles remain well-defined. In the dirty regime, transport is dominated by extrinsic mechanisms such as skew scattering and side jump, which are not captured in the present framework. Also, the equations are preferably suited for the low-temperature regime. In the higher temperature limit (such as 300 $K$), scattering mechanisms starts affecting the experimental data, which is not captured in the discussed equations. Thus, one should consider these aspects while comparing the obtained results with the experimental data. It is important to highlight here that theoretical studies have been previously carried out for room temperature using the Eqs. \ref{eqAHE} and \ref{eqANE}\cite{yao2004first,li2023anomalous,guo2014anomalous}.

Presently, the most common approach to calculate AHC or ANC is based on tight-binding (TB) model. Such TB models are usually generated using wannierization techniques as implemented in Wannier90 package\cite{mostofi2008wannier90}. In this method, maximally localized Wannier functions (MLWFs) are constructed from the Bloch states obtained in density functional theory (DFT) calculations\cite{marzari2012maximally,marrazzo2024wannier}. These MLWFs serve as basis set, in terms of which, Bloch states (\textit{n},\textbf{\textit{k}}) corresponding to any band index \textit{n} \& crystal momentum \textit{\textbf{\textit{k}}} is reproduced using wannier interpolation techniques. Using these reproduced Bloch states, the energy eigenvalues and momentum-matrix elements are calculated, which are further used to compute $\Omega_{\mu\nu}^n(\textbf{\textit{k}})$ \& AHC/ANC. This method is implemented in packages like Wannier90\cite{mostofi2008wannier90} and WannierTools\cite{wu2018wanniertools}. It is important to note here that the accuracy of calculated AHC/ANC depends on the quality of the Wannier functions obtained via wannierization technique. For the case of AHC/ANC computation, the high quality wannier functions (HQWFs) are one that perfectly reproduce the Bloch states with correct energy and associated momentum-matrix elements in the energy-region of interest. Moreover, the number of these wannier functions must be as small as possible. However, to obtain HQWFs, the DFT bands in the required energy window must be disentangled from the bands outside this window. The window is popularly known as disentangled energy window. Additionally, the projectors contributing to the bands in disentangled energy window must have negligible contribution to other bands outside the window. Another general observation is that obtaining HQWFs becomes extremely difficult when the bands in the region of interest are greatly entangled among themselves and are highly dispersive. Apart from these, wannierization procedure is highly parameter dependent and there is no straight-forward procedure to obtain HQWFs\cite{pandey2023py}. Moving to the case of real materials, there may be the case where the dispersion curve may not possess disentangled energy window. Furthermore, the bands may be extremely dispersive as in case of metals. In such situations, obtaining HQWFs is a hopeless situation and thus the reliability of AHC/ANC obtained from the wannier interpolation becomes questionable. 

An alternative to Wannier-based methods is the direct calculation of $\boldsymbol\Omega$ from \textit{first-principles} wave functions. This approach avoids interpolation errors by working directly with the original DFT eigenfunctions and momentum-matrix elements, thereby eliminating the need for wannier functions. Such implementation has been previously carried out to compute AHC in the LmtART package, which uses full-potential linear muffin-tin orbital (FP-LMTO) basis set\cite{savrasov2005program}. In contrast to LmtART package, WIEN2k package uses full-potential linearized augmented-plane-wave (FP-LAPW) basis set\cite{blaha2020wien2k}. In the present DFT community, FP-LAPW basis set is claimed to be the most accurate one\cite{wimmer1981full,jansen1984total,mattheiss1986linear,BLAHA1990399}. Hence, AHC or ANC computed from a code that directly uses eigenvalues and momentum-matrix elements obtained from WIEN2k calculations is expected to be more accurate and reliable. Additionally, WIEN2k is one of the most widely used DFT package. Thus, implementing the calculation of the discussed transverse conductivities using the output of WIEN2k calculations will be beneficial for vast research audience. Now we shall discuss the aspects that can enhance the efficiency of code for such implementation.

The transport properties such as AHC \& ANC generally get converged for \textbf{\textit{k}}-mesh size of 400$\times$400$\times$400 or more. Moreover, if the bands\textquoteright \hspace*{0.02in} resolved $\boldsymbol\Omega$ at each \textit{\textbf{\textit{k}}}-points is stored, then AHC/ANC can be easily calculated at any value of $\omega$ and $T$ in the form of post-process. In such a scenario, parallel computing and efficient handling of data will greatly enhance the efficiency of the code. By efficient handling of data, we mean that one which occupy less memory and is easily accessible. This can be achieved by storing the coordinates of \textit{\textbf{\textit{k}}}-points and the values of bands\textquoteright \hspace*{0.02in} resolved $\boldsymbol\Omega$ in a binary file. The binary file is not directly accessible. Thus, one can only store important information without any extra formatting or comment lines. Furthermore, storing data in binary format greatly reduces the size of the file (about 10-15 times). Moving to parallel computing, for the present case the calculation of $\boldsymbol\Omega$ can be made parallel over \textbf{\textit{k}}-points. In addition to this, the calculation of integrand in equations \ref{eqAHE} and \ref{eqANE} can also be made parallel over \textbf{\textit{k}}-points. Another aspect is that previous study revealed that parallelism in C++ (using \textit{OpenMP}) outperforms python (using \textit{multiprocessing} module)\cite{arboleda2023performance}. Furthermore, compiling C++ modules produce executable files (in machine level language) which performs a task much faster as compared to doing same task using languages like python. Designing a \textit{first-principles} code to compute AHC/ANC by considering all these aspects is expected to be highly efficient and less time consuming.

In this work, we present a C++ based code named \textit{C-BerryTrans} to compute $\boldsymbol\Omega$ and related AHC \& ANC directly from WIEN2k output. The code extracts eigenvalues and momentum-matrix elements from WIEN2k computations and evaluates $\boldsymbol\Omega$ explicitly by invoking the Kubo-like formula\cite{thouless1982quantized}. For efficient handling of large number of \textit{\textbf{\textit{k}}}-points, calculations of $\boldsymbol\Omega$ are parallelized over \textit{\textbf{\textit{k}}}-points. Furthermore, unlike in WannierTools where calculations of $\boldsymbol\Omega$ and AHC/ANC is done corresponding to a single value of $T$ at a time, \textit{C-BerryTrans} first calculates band-resolved $\boldsymbol\Omega$ and stores it in a binary file (thereby reducing the memory required for storage). Then, based on the provided window of $T$ and $\omega$ in the input file, it computes AHC/ANC as post-process. While the majority of computation time is involved in calculations of $\boldsymbol\Omega$, the post-process is comparatively much faster. Thus, \textit{C-BerryTrans} can be used to efficiently compute AHC/ANC for a range of $\omega$ and $T$ values in a single run once the $\boldsymbol\Omega$ file is obtained. Additionally, \textit{C-BerryTrans} provides a direct and interpolation-free alternative, ensuring numerical accuracy and scalability. This is particularly advantageous for high-throughput materials discovery, where manual fine-tuning of Wannier functions is impractical. The code is also provided with the option to calculate $\boldsymbol\Omega$ along user-defined \textbf{\textit{k}}-point paths with Fermi-Dirac or entropy weighting for AHC and ANC, respectively. This provides valuable insight into the connection between band crossings, dispersion features and topological responses. To validate the code, it has been tested on some well-known materials exhibiting AHC (Fe, Fe$_3$Ge \& Co$_2$FeAl) and  ANC (Pd, Fe$_3$Al \& Co$_2$FeAl). The results obtained using the \textit{C-BerryTrans} code are compared with the values reported in literature corresponding to each material. Additionally, the code includes a visualization module, \textit{berryTrans\_plot.py}, which provides interactive options to visualize and analyse the distribution of $\Omega_{\mu \nu}(\textbf{\textit{k}})*FermiDistributionFunction$=$\sum_n f_n(\textbf{\textit{k}}) \Omega_{\mu \nu}^n(\textbf{\textit{k}})$ or $\Omega_{\mu \nu}(\textbf{\textit{k}})*EntropyDistributionFunction$=\\$\sum_n \Omega_{\mu \nu}^n(\textbf{\textit{k}}) \left[ (E_n - E_F) f_n(\textbf{\textit{k}}) + k_B T \ln \left(1 + \exp\left( \frac{E_n - E_F}{-k_B T} \right) \right) \right]$ in the \textbf{\textit{k}}-space. Here, $\mu$, $\nu$=$x$, $y$ or $z$. This further enhances the capability of the code in exploring topological materials.


\section{Theoretical Background} \label{TheoreticalBackground}

\subsection{Berry curvature and anomalous transverse conductivities}
As previously discussed, AHC and ANC originates from the $\boldsymbol\Omega$ of electronic bands in momentum space. \textit{C-BerryTrans} directly calculates the antisymmetric tensor $\Omega_{\mu \nu}^n$ from the WIEN2k output by invoking the Kubo-like formula\cite{stejskal2023theoretical,thouless1982quantized,nagaosa2010anomalous,xiao2010berry,ernst2019anomalous}:
\begin{equation}
   \Omega_{\mu \nu}^n(\textbf{\textit{k}})=-\frac{\hbar^2}{m^2} \sum_{n^{\prime} \neq n} \frac{2 \operatorname{Im}\left\langle\psi_{n \textbf{\textit{k}}}\right| p_\mu\left|\psi_{n^{\prime} \textbf{\textit{k}}}\right\rangle\left\langle\psi_{n^{\prime} \textbf{\textit{k}}}\right| p_\nu\left|\psi_{n \textbf{\textit{k}}}\right\rangle}{\left(E_n-E_{n^{\prime}}\right)^2}
   \label{eq2}
\end{equation}
where $E_n$ is the band-energy, $\psi_{n \textbf{\textit{k}}}$ are the Bloch states and $p_\mu$ are the momentum operators. In addition to this, $\hbar$ \& $e$ denote the reduced Planck\textquoteright s constant and electronic charge, respectively. Here, $\mu$, $\nu$ \& $\xi$ represent $x$, $y$ \& $z$ cartesian direction as per required by the expression. Using $\Omega_{\mu \nu}^n$, $\sigma_{\mu \nu}^{\mathrm{AHC}}$ is calculated by using Eq. \ref{eqAHE} whereas $\alpha_{\mu \nu}^{\mathrm{ANC}}$ is estimated via Eq. \ref{eqANE}. It is important to note that Eqs. \ref{eqAHE} \& \ref{eqANE} contain $f_n(\textbf{\textit{k}})$ \& $T$. This allows the computation of AHC \& ANC at various combinations of $\omega$ and $T$ using the $\Omega_{\mu \nu}^n$ values.

An important aspect to highlight here is the role of spin-orbit coupling (SOC) in determining the above discussed transverse responses. If SOC strength in a system is zero, then the up-spin and down-spin sub systems are independent and each is governed by a real Hamiltonian. In such a scenario, $\Omega_{\mu \nu}^n(\textbf{\textit{k}})$=$-\Omega_{\mu \nu}^n(-\textbf{\textit{k}})$ for all combinations of $n$ \& \textbf{\textit{k}}. Thus, the integrals in Eqs. \ref{eqAHE} \& \ref{eqANE} vanish. This situation arises because the spontaneously broken time-reversal symmetry in the spin channel is not communicated to the orbital channel, which is responsible for giving rise to transverse responses\cite{vanderbilt2018}. Thus for computing AHC or ANC, SOC-effect must be necessarily taken into account. \textit{C-BerryTrans} is designed considering this aspect.

It is also important to note that Eq. \ref{eq2} formally exhibits singular behavior when two bands become degenerate at a given \textbf{\textit{k}}-point (such as Weyl nodes) due to the vanishing energy denominator. However, such divergences cancel out in the evaluation of physical observables such as AHC and ANC. Specifically, for a pair of degenerate bands (lets say, with band-indices $n$ \& $m$), the $\boldsymbol\Omega$ contributions are equal in magnitude and opposite in sign. Furthermore, at the degenerate point, the value of $f(\textbf{\textit{k}})$ (and hence the entropy function) is the same for bands with indices $n$ \& $m$. Thus, their contributions to AHC or ANC cancel out at the singularity points. In the present implementation, bands with energy differences below 1E-8 Rydberg are treated as degenerate, and their contributions are excluded from the $\boldsymbol\Omega$ calculation. This procedure avoids numerical instability while preserving the accuracy of the integrated transport coefficients.

\subsection{Extracting Berry curvature from WIEN2k output}
WIEN2k is a full-potential linearized augmented plane wave (FP-LAPW) method-based DFT package\cite{blaha2020wien2k}. The momentum-matrix elements required for $\boldsymbol\Omega$ calculations can be obtained using WIEN2k's OPTIC module, which computes: $\textless \psi_{n\textbf{\textit{k}}}| \textbf{\textit{p}}| \psi_{m\textbf{\textit{k}}} \textgreater$, where $\textbf{\textit{p}}$ is the momentum operator\cite{WIEN2kUserManual}. The workflow for extracting $\boldsymbol\Omega$ involves- (I) Perform self-consistent SOC included DFT calculations to obtain ground state charge density, (II) obtain the eigenvalues at a given \textbf{\textit{k}}-point, (III) obtain the momentum-matrix elements at the given \textbf{\textit{k}}-point (found in \textit{case.pmat} file), and (IV) calculate $\Omega_{\mu \nu}^n(\textbf{\textit{k}})$ at the given \textbf{\textit{k}}-point using Eq. \ref{eq2}. In the present DFT community, FP-LAPW basis set is considered as the most accurate one. Thus, the obtained values of $\Omega_{\mu \nu}^n(\textbf{\textit{k}})$ are expected to be highly accurate.

\subsection{Numerical integration over the entire BZ}
As seen from Eqs. \ref{eqAHE} \& \ref{eqANE}, both $\sigma_{\mu\nu}^{\text{AHC}}$ and $\alpha_{\mu\nu}^{\text{ANC}}$ involve BZ integration of the components of $\boldsymbol\Omega$, weighted respectively by the $f_n(\textbf{\textit{k}})$ and the entropy-related function. In \textit{C-BerryTrans}, these integrals are approximated by discrete summations over bands and $\textbf{\textit{k}}$-points: specifically, sums of the form $\sum_{n,\textbf{\textit{k}}} f_n(\textbf{\textit{k}})\, \Omega_{\mu\nu}^{n}(\textbf{\textit{k}})$ for AHC, and $\sum_{n,\textbf{\textit{k}}} \Omega_{\mu\nu}^{n}(\textbf{\textit{k}})$ multiplied by the entropy kernel for ANC, evaluated over a dense, uniformly distributed $\textbf{\textit{k}}$-mesh.


\begin{table*}
\caption{\label{tab71}
{Descriptions of the various input parameters for the \textit{C-BerryTrans} code.}}
 \centering
 \small 
 \begin{tabular}{|c|c|}
   \hline
    {\textbf{Name}} &{\textbf{Meaning}} \\
   \hline
    
\textit{case} & Name of the WIEN2k directory containing self-consistently converged results with SOC included.\\\hline
\textit{struct\_num} & Numerical identifier for the crystal structure type (see table \ref{tab72} for reference). \\ \hline
\textit{chemical\_potential} & Central value of the $\omega$ (in Rydberg) used to define the energy window for AHC/ANC calculation.\\\hline
\textit{emax} & Upper bound of the $\omega$ window (in meV) relative to the specified \textit{chemical\_potential}. \\ \hline
\textit{emin} & Lower bound of the $\omega$ window (in meV) relative to the specified \textit{chemical\_potential}. \\ \hline
\textit{estep} & Increment (in meV) for varying the $\omega$ value within the defined energy window. \\ \hline
\textit{Tmin} & Minimum value of $T$ (in Kelvin) at which AHC/ANC will be evaluated. \\  \hline
\textit{Tmax} & Maximum value of $T$ (in Kelvin) at which AHC/ANC will be evaluated. \\  \hline
\textit{Tstep} & Step size (in Kelvin) for varying $T$ between \textit{Tmin} and \textit{Tmax}. \\ \hline
\textit{spin\_pol} (1/0) & Indicates whether spin-polarization is considered: 1 for spin-polarized systems, 0 otherwise. SOC is\\
                        & always included in \textit{C-BerryTrans} calculations. \\ \hline
\textit{kmesh} & Size of \textbf{\textit{k}}-mesh to be produced at which $\boldsymbol\Omega$ must be computed. \\ \hline
\textit{shift\_in\_k} (1/0) & Shift of \textbf{\textit{k}}-mesh from high-symmetric points is needed (1) or not (0). \\ \hline
\textit{NUM\_THREADS} & Number of threads over which computation of $\boldsymbol\Omega$ and transport (AHC \& ANC)  over \textbf{\textit{k}}-points must\\
                      & be made parallel. \\ \hline
\textit{num\_kp\_at\_once} & Number of \textbf{\textit{k}}-points processed in a single batch per thread to manage RAM usage efficiently and\\
                           & avoid overflow. \\ \hline

\textit{calc\_energy\_min} & Lower energy limit (in meV and relative to the specified \textit{chemical\_potential}) up to which eigenvalues\\
                           & and momentum-matrix elements must be computed.\\ \hline
\textit{calc\_energy\_max} & Upper energy limit (in meV and relative to the specified \textit{chemical\_potential}) up to which eigenvalues\\
                           & and momentum-matrix elements must be computed.\\ \hline
\textit{calcEnSuitCheck}   & 1 (highly recommended)  $\rightarrow$ checks if the provided value of \textit{calc\_energy\_min} \& \textit{calc\_energy\_max} will\\
                           & not result in \textit{TAPEWF} errors, 0 $\rightarrow$ avoids this check. WARNING: Too small value of \textit{calc\_energy} \\
                           & window may result is several \textit{TAPEWF} and \textit{segmentation fault (core dumped)} errors.\\\hline     
\textit{property}  & Which property must be computed in the post-process: 1$\rightarrow$ AHC, 2$\rightarrow$ ANC.\\ \hline
\textit{band}  & 1 $\rightarrow$ $\boldsymbol{\Omega}\!\times\! f_n(\textbf{\textit{k}})$ (AHC) or $\boldsymbol{\Omega}\!\times\! entropy$ (ANC) along \textbf{\textit{k}}-path; 
0 $\rightarrow$ standard AHC/ANC computation.\\ \hline
\textit{berryTrans\_plot\_at\_T}  & Value of $T$ (in Kelvin) at which $\Omega_{\mu \nu}(\textbf{\textit{k}})*FermiDistributionFunction$=$\sum_n f_n(\textbf{\textit{k}}) \Omega_{\mu \nu}^n(\textbf{\textit{k}})$\\
                              &  (when $property$=1) or $\Omega_{\mu \nu}(\textbf{\textit{k}})*EntropyDistributionFunction$=\\
                              & $\sum_n\Omega_{\mu \nu}^n(\textbf{\textit{k}}) \left[ (E_n - E_F) f_n(\textbf{\textit{k}}) + k_B T \ln \left( 1 + \exp\!\left( \frac{E_n - E_F}{-k_B T} \right) \right) \right]
$ (when $property$=2) must be plotted\\
&  using the \textit{berryTrans\_plot.py} module of \textit{C-BerryTrans} code. Here, $\mu$, $\nu$=$x$, $y$ or $z$.\\
    
   \hline
 \end{tabular}
 \label{tab71}
\end{table*}

\begin{table}
\caption{\label{tab72}
{The structure number assigned to different crystal structures.}}
 \centering
 \small 
 \begin{tabular}{|c|c|}
   \hline
    {\textbf{Crystal Structure}} &{\textbf{Structure Number}} \\
   \hline
          Cubic Primitive           & 1\\
          Cubic face-centred        & 2\\
          Cubic body-centred        & 3\\
          Tetragonal Primitive      & 4\\
          Tetragonal body-centred   & 5\\
          Hexagonal Primitive       & 6\\
          Orthorhombic Primitive    & 7\\
          Orthorhombic base-centred & 8\\
          Orthorhombic body-centred & 9\\
          Orthorhombic face-centred & 10\\
          Trigonal Primitive        & 11 \\

   \hline
 \end{tabular}
 \label{tab72}
\end{table}

\section{Workﬂow}
\label{C-BerryTransDiscussion}

In order to use the \textit{C-BerryTrans} code, user needs to extract the zipped file of \textit{C-BerryTrans} code. This will produce a \textit{C-BerryTrans} directory containing sixteen C++ modules along with the \textit{Makefile}, \textit{C-BerryTrans.input} file and a python module named \textit{berryTrans\_plot.py}. Now, user must install the code by running the \textit{make} command inside \textit{C-BerryTrans} directory. This will produce the executable files corresponding to the sixteen C++ modules of the \textit{C-BerryTrans} code. Finally, user is required to define the path of the \textit{C-BerryTrans} folder in \textit{.bashrc} file. With this, the installation procedure is complete. The detailed working of the code is described below.

For running the intrinsic transverse response (AHC or ANC) calculations, one needs to first obtain the \textit{case} folder containing SOC included self-consistently converged ground state files corresponding to the given material from WIEN2k package. It is important to note that the self-consistently converged ground state calculations can be done over the \textbf{\textit{k}}-points sampled in the irreducible parts of the BZ (IBZ). The obtained ground state converged files enable the computation of eigenvalues and momentum-matrix elements at any arbitrary \textbf{\textit{k}}-point. Nextly, \textit{C-BerryTrans.input} file must be prepared. All the necessary input variables corresponding to intrinsic transverse response calculations must be defined in it. These input variables are briefly described in table \ref{tab71}. The name of folder containing self-consistently converged ground-state files must be assigned to \textit{case} variable. The structure number of material under study must be given to \textit{struct\_num} variable. In this regard, table \ref{tab72} defines the structure number assigned to various crystal structures in \textit{C-BerryTrans} code. Furthermore, it is essential to specify whether the material under investigation is spin-polarized by assigning a value of 1 (for spin-polarized case) or 0 (for non-spin-polarized case) to the \textit{spin\_pol} variable. One must note here that SOC is always taken into account in the \textit{C-BerryTrans} calculations. Moving next, the momentum-matrix elements, in general, do not follow the same symmetry relations as the eigenvalues. As a result, it is not straightforward to employ symmetry operations for the calculation of AHC/ANC. Therefore, \textit{C-BerryTrans} performs the $\boldsymbol\Omega$ calculation over all \textbf{\textit{k}}-points in the full BZ. In this regard, it is commonly observed that WIEN2k takes exceptionally long time to generate the \textit{case.klist} file for dense \textbf{\textit{k}}-meshes (for instance, $400 \times 400 \times 400$). To overcome this bottleneck, \textit{C-BerryTrans} provides a dedicated module named \textit{klist\_gen\_div} to efficiently generate the required \textbf{\textit{k}}-list file corresponding to the full BZ. In this regard, if one wishes to calculate AHC/ANC on a $400 \times 400 \times 400$ \textbf{\textit{k}}-mesh, the string \textquoteleft 400 400 400\textquoteright\hspace*{0.02in} must be assigned to the \textit{kmesh} variable in the \textit{C-BerryTrans.input} file. Using this information, along with the value of the \textit{struct\_num} variable, the \textit{klist\_gen\_div} module efficiently produces the \textit{case.klist} file in the full BZ. To obtain a shifted \textbf{\textit{k}}-mesh (ensuring that none of the \textbf{\textit{k}}-points coincide with high-symmetry points), one must set the variable \textit{shift\_in\_k} to 1; otherwise, it should be set to 0. In \textit{C-BerryTrans}, the \textbf{\textit{k}}-points are generated by first scaling the BZ in terms of the primitive reciprocal lattice vectors ($g_{1}, g_{2}, g_{3}$) within the range 0-1. If the \textbf{\textit{k}}-mesh of size $n_{1} \times n_{2} \times n_{3}$ is needed, then the directions along $g_{1}$, $g_{2}$ and $g_{3}$ are subdivided into $n_{1}, n_{2},$ and $n_{3}$ parts, respectively, to form a uniform grid in primitive reciprocal space. Since the primitive reciprocal vectors are defined with respect to the cartesian axes ($k_{x}, k_{y}, k_{z}$), the grid points are subsequently transformed into cartesian coordinates. For this conversion, \textit{WIEN2k} adopts the Cracknell convention, and the same has been implemented in the \textit{C-BerryTrans} code to ensure full compatibility\cite{bradley2009mathematical}. Finally, the \textbf{\textit{k}}-point coordinates are written in the \textit{case.klist} format required by \textit{WIEN2k}, enabling seamless integration with its workflow. The \textit{case.klist} file generated by \textit{klist\_gen\_div} is identical to the one produced by WIEN2k in the full BZ for the specified \textbf{\textit{k}}-mesh size. One must note here that, in the present version of the \textit{C-BerryTrans} code, \textit{klist\_gen\_div} module does not produce the \textbf{\textit{k}}-mesh in the IBZ. Moving further, the calculation of $\boldsymbol\Omega$, AHC \& ANC via \textit{C-BerryTrans} can be performed in serial or parallel mode depending on whether the value assigned to \textit{NUM\_THREADS} variable is 1 or greater than 1, respectively. If \textit{NUM\_THREADS}$>$1, then \textit{klist\_gen\_div} divides the generated \textit{case.klist} file into \textit{NUM\_THREADS} number of files with names \textit{case.klist1}, \textit{case.klist2}, ..., \textit{case.klistNUM\_THREADS}. The total number of \textbf{\textit{k}}-points in \textit{case.klist} file is equally distributed among these files. However, if some \textbf{\textit{k}}-points remain extra after this equal distribution, they are adjusted in the last \textit{case.klistX} file. Nextly, \textit{C-BerryTrans} can compute AHC or ANC for a range of $T$ values in a given $T$ window at a time. For this, user needs to provide the lower-limit, upper-limit and the temperature step to variable \textit{Tmin}, \textit{Tmax} and \textit{Tstep}, respectively. All these values must be given in units of Kelvin. At the same time, for each $T$ value, AHC or ANC can be computed for a range of $\omega$ values within a given energy window range. For this, user needs to firstly assign the $\omega$ value (in Rydberg) to the variable \textit{chemical\_potential}. Then, the lower-range (in meV and with respect of the value assigned to \textit{chemical\_potential}), upper-range (in meV and with respect of the value assigned to \textit{chemical\_potential}) and step-size of $\omega$ (in meV) must be assigned to variables \textit{emin}, \textit{emax} and \textit{estep}, respectively. Moving further, it is important to highlight here that while calculating the eigenvalues via WIEN2k, the memory occupied in RAM increases with the increase in number of \textbf{\textit{k}}-points in \textit{case.klist} file. Thus, depending on the size of RAM, beyond a certain size of \textbf{\textit{k}}-mesh, RAM gets fully occupied and the calculations get aborted. To avoid this situation, \textit{C-BerryTrans.input} file is provided with a variable named \textit{num\_kp\_at\_once}. The value of this variable decides the number of \textbf{\textit{k}}-points over which $\boldsymbol\Omega$ is calculated at each thread at any given time. The strategy followed is that the original \textit{case.klistX} file at each thread is renamed as \textit{case.klist1}. Now, based on the value assigned to $num\_kp\_at\_once$, that number of \textbf{\textit{k}}-points will be extracted from the \textit{case.klist1} file, and a new \textit{case.klist} file will be created. Using this file, eigenvalues and momentum-matrix elements are calculated via WIEN2k following which $\boldsymbol\Omega$ is calculated and stored. Having this done, again a new \textit{case.klist} file is created with next $num\_kp\_at\_once$ number of \textbf{\textit{k}}-points from \textit{case.klist1} file and the process is repeated until all the \textbf{\textit{k}}-points are processed. Moving next, \textit{C-BerryTrans} can operate in two different computational modes depending on the value assigned to the \textit{band} variable in the \textit{C-BerryTrans.input} file. If \textit{band=0}, the code performs the conventional AHC/ANC calculation, where the $\boldsymbol{\Omega}$ is computed over a dense \textit{\textbf{k}}-mesh covering the full BZ and subsequently integrated to obtain the transverse transport coefficients. On the other hand, assigning \textit{band=1} activates the \textit{\textbf{k}}-path-resolved mode. In this case, instead of generating a dense full-BZ \textit{\textbf{k}}-mesh, the code evaluates the $\boldsymbol{\Omega}$-weighted AHC/ANC contributions along a user-defined high-symmetry \textit{\textbf{k}}-path (see section \ref{lab1}). This mode is particularly useful for analysing how specific band crossings and dispersion features contribute to the anomalous transverse responses. Lastly, as already described, ANC depends only on the states around the Fermi energy. Also, a close inspection of Eqs. \ref{eqAHE} \& \ref{eq2} suggest that for a given value of $T$, a pair of bands will not contribute to AHC if at each \textbf{\textit{k}}-points of the BZ, the occupancy of one band is exactly equal to the other band. The pair of bands will only have non-zero contribution to total AHC at the given value of $T$, if in some regions of the BZ, the occupancy of one of the band differs from the other\cite{wang2006ab}. Next thing that these equations suggest is that if a pair of bands have large energy difference, then their contributions to $\boldsymbol\Omega$ and hence to transverse responses will be negligible. Thus, contributions from such pairs can be safely ignored in computations. This discussion suggests that calculations of all the bands at a given \textbf{\textit{k}}-point, which generally consume significant computational resources, are not always required. Since the computational time strongly depends on the number of bands for which eigenvalues and momentum-matrix elements are evaluated, it is advantageous to choose an energy window that is sufficiently narrow while still containing the relevant bands contributing to AHC/ANC. Therefore, for temperatures up to 300 K, the AHC/ANC at a given $\omega$ can generally be computed by considering an energy interval of not more than $\pm$1.5 eV with respect to $\omega$. In \textit{C-BerryTrans}, the user can specify an energy window within which the eigenvalues and momentum-matrix elements are calculated. The lower and upper limits of this window (in meV and relative to the value provided to the \textit{chemical\_potential} variable) are assigned to the variables \textit{calc\_energy\_min} and \textit{calc\_energy\_max} in the input file, respectively. Based on these values, \textit{C-BerryTrans} automatically modifies the \textit{case.in1}, \textit{case.inso}, and \textit{case.inop} files required for the WIEN2k calculations, thereby improving computational efficiency. The details of the structure of these files can be found in the WIEN2k userguide\cite{WIEN2kUserManual}. It is important to mention here that, the chosen energy window should not be excessively small and must ensure that at least one band exists within the specified range at every \textbf{\textit{k}}-point sampled across the BZ. Otherwise, WIEN2k may encounter \textit{TAPEWF} errors during the eigenvalue and momentum-matrix-element calculations. Such errors will lead to unphysical results and hence, one must abort the calculation if it appears. A practical procedure to determine an appropriate energy window is to first perform a density of states (DOS) calculation and identify an energy range around the Fermi level (value assigned to \textit{chemical\_potential} variable) where electronic states are present, ensuring that the selected window does not lie entirely inside a band gap, if present. Subsequently, one should compute the band structure along high-symmetry \textbf{\textit{k}}-paths and verify that the chosen energy window includes at least one band at every \textbf{\textit{k}}-point along these paths. To further assist users, the \textit{C-BerryTrans} code includes an automated pre-check (using \textit{calcEnSuitCheck} module) before starting the $\boldsymbol\Omega$ calculations. The code verifies whether the specified energy window is sufficiently large to avoid \textit{STOP TAPEWF} errors. This checking step generally requires only a few minutes, depending on the material system. If the selected window (based on the values provided to the variables \textit{calc\_energy\_min} and \textit{calc\_energy\_max}) is found to be inadequate, the code issues the warning
\begin{center}
\textit{WARNING: calc\_energy\_min and calc\_energy\_max window must be larger to avoid TAPEWF errors}
\end{center}
and safely terminates the calculation. Therefore, users are advised to begin with a reasonably small but physically meaningful energy window (for example, $\pm$0.5 eV around the \textit{chemical\_potential}) and gradually increase the window size if the above warning appears, until the calculation proceeds without errors. It is important to note that once a suitable energy window has been identified for a given material system, users may need to restart the calculations multiple times due to unrelated computational or technical reasons (for instance, changing the values of variables \textit{NUM\_THREADS} or \textit{num\_kp\_at\_once}). In such situations, repeating the automated pre-check unnecessarily increases the total computational time. Therefore, \textit{C-BerryTrans} provides the variable \textit{calcEnSuitCheck} in the input file. Setting \textit{calcEnSuitCheck=1} activates the energy-window suitability check, whereas assigning \textit{calcEnSuitCheck=0} skips this verification step. This strategy ensures both numerical stability and computational efficiency while maintaining the physical relevance of the calculated AHC/ANC values. Now, a brief discussion about the workflow of the code is mentioned below. 

\begin{figure*}[t]
    \centering
    \includegraphics[width=0.65\textwidth,height=14.0cm]{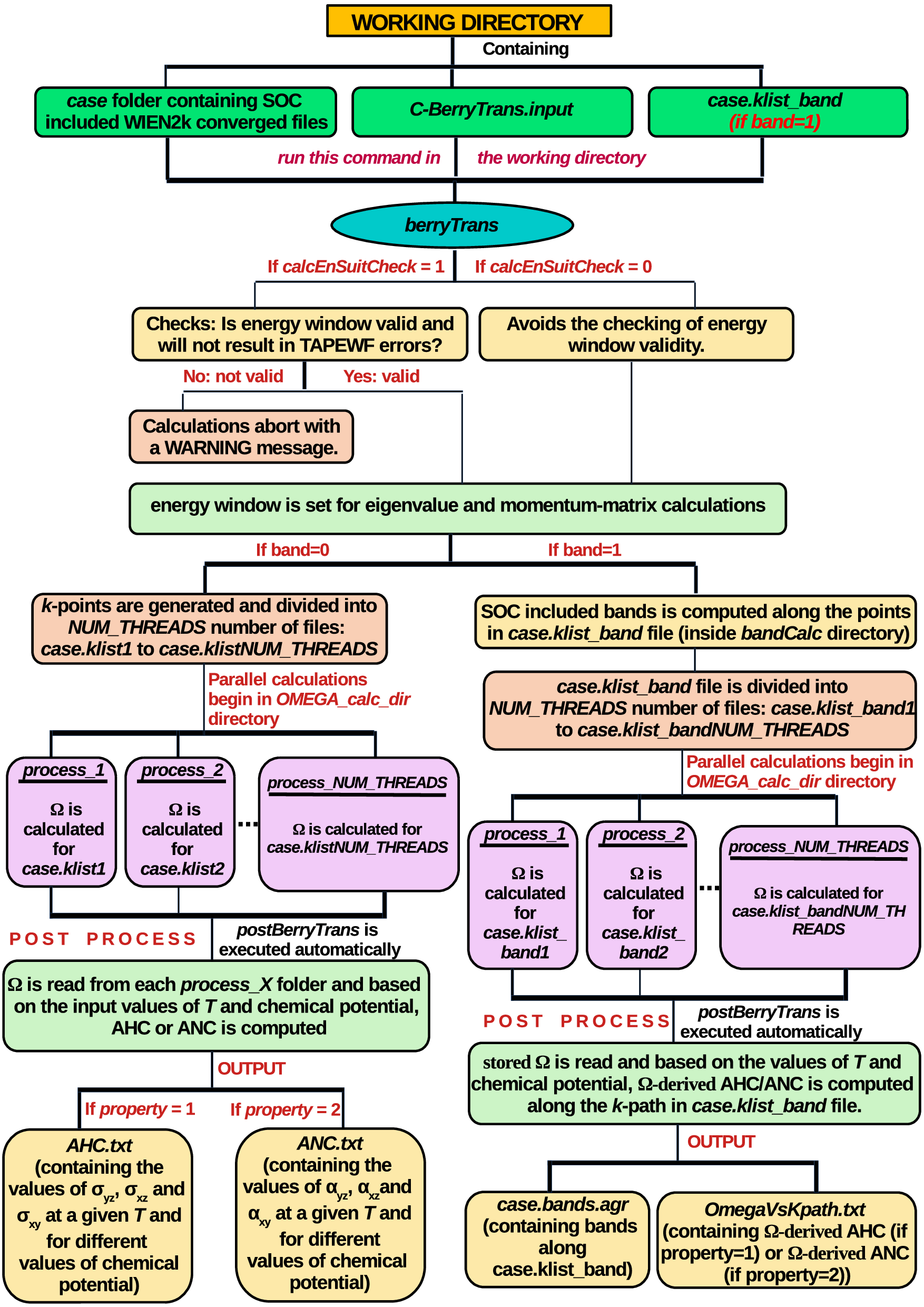}
    \caption{Workflow of the \textit{C-BerryTrans} code showing the sequence of input preparation, $\boldsymbol\Omega$ calculation, and post-processing.}
    \label{workflow}
\end{figure*}

The flowchart of the working of \textit{C-BerryTrans} code is shown in Fig. \ref{workflow}. As already discussed, the code supports two different calculation modes depending on the value assigned to the \textit{band} variable. The workflow corresponding to the case when \textit{band=0} is discussed now. For the other case (\textit{i.e.}, when \textit{band=1}), the working is almost similar and will be discussed later in section \ref{lab1}. The calculation of band-resolved $\boldsymbol\Omega$ can be started by running the command: \textit{berryTrans}. Before proceeding to the main calculations, the code has an option to first perform a pre-check of the user-defined calculation energy window (controlled by the variable \textit{calcEnSuitCheck}). If \textit{calcEnSuitCheck} is set to 1, the module verifies whether the chosen energy window defined by the values assigned to \textit{calc\_energy\_min} and \textit{calc\_energy\_max} is sufficient to avoid \textit{TAPEWF} errors. In case the window is found inadequate, the code issues a warning and terminates safely. If the check is satisfied, or if \textit{calcEnSuitCheck} is set to 0 (thereby skipping the verification), the code proceeds to the next stage, where it invokes the \textit{modify\_calc\_window} module to generate the \textit{case.inop} file. After this, the module modifies the energy window in \textit{case.in1}, \textit{case.inso} and \textit{case.inop} files as per the values provided to the variables \textit{calc\_energy\_min} \& \textit{calc\_energy\_max} in the input file. Having this done, the code executes the \textit{klist\_gen\_div} module to generate the \textit{case.klist} file based on the \textbf{\textit{k}}-mesh size assigned to \textit{kmesh} variable. Nextly, if \textit{NUM\_THREADS}$>$1, the \textbf{\textit{k}}-points in \textit{case.klist} file is equally divided into \textit{NUM\_THREADS} number of files as described above. Now, a directory named \textit{OMEGA\_calc\_dir} is created within which \textit{NUM\_THREADS} number of subdirectories are created with names- \textit{process\_1}, \textit{process\_2}, ..., \textit{process\_NUM\_THREADS}. In each of these subdirectories, the \textit{case} folder along with \textit{case.inop} and \textit{case.klistN} (where \textit{N} is the number associated with the subdirectory) files are copied. The \textit{case.klistN} file is renamed as \textit{case.klist1} after being copied in the subdirectories. Now, the calculation of eigenvalues and momentum-matrix elements along with the computation of $\boldsymbol\Omega$ corresponding to \textit{case.klist1} file in each of these subdirectories will be carried out at separate threads. In this way the computation of $\boldsymbol\Omega$ is made parallel. As highlighted above, to avoid the over usage of RAM, in each of the subdirectories, firstly \textit{num\_kp\_at\_once} number of \textbf{\textit{k}}-points will be taken out of \textit{case.klist1} file to make a new file named \textit{case.klist}. Now the calculation of eigen energy and momentum-matrix elements corresponding to the generated \textit{case.klist} file will be done. After this, the band-resolved $\boldsymbol\Omega$ will be calculated for the \textbf{\textit{k}}-points in \textit{case.klist} file by calling the module \textit{energy\_pmat\_read}. The obtained result will be stored in binary file named \textit{data.bin}. Again, the next \textit{num\_kp\_at\_once} number of \textbf{\textit{k}}-points will be taken out of \textit{case.klist1} and the process will be repeated till the computation of band-resolved $\boldsymbol\Omega$ is done for all the \textbf{\textit{k}}-points in \textit{case.klist1} file. After this process is complete, everything is set for the post-process (calculation of $\sigma_{\mu \nu}^{\mathrm{AHC}}$ or $\alpha_{\mu \nu}^{\mathrm{ANC}}$).\\\\

\begin{figure*}[t]
    \centering
    \begin{subfigure}[b]{0.35\textwidth}
        \includegraphics[width=0.80\textwidth,height=4.0cm]{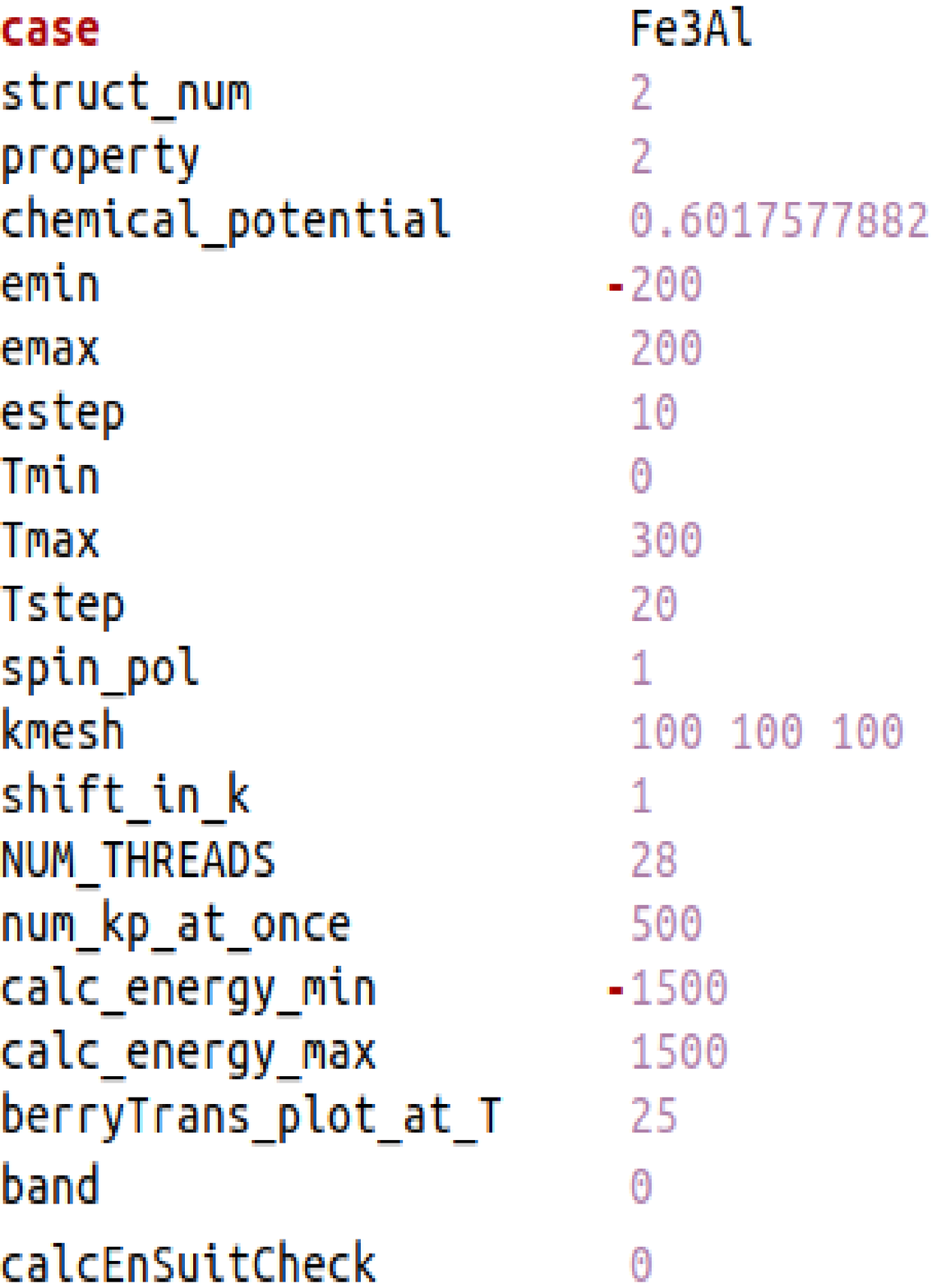}
        \caption{\textit{C-BerryTrans.input} file}
        \label{klist}
    \end{subfigure}
    \begin{subfigure}[b]{0.55\textwidth}
        \includegraphics[width=1.00\textwidth,height=4.0cm]{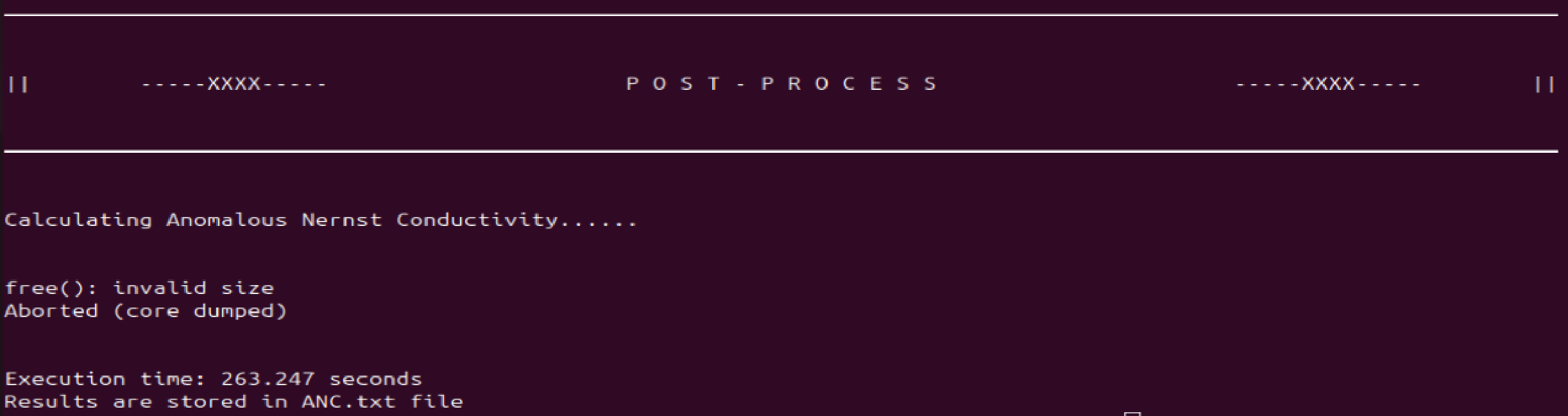}
        \caption{Output}
        \label{energy}
    \end{subfigure}
    \begin{subfigure}[b]{0.35\textwidth}
        \includegraphics[width=0.80\textwidth,height=4.0cm]{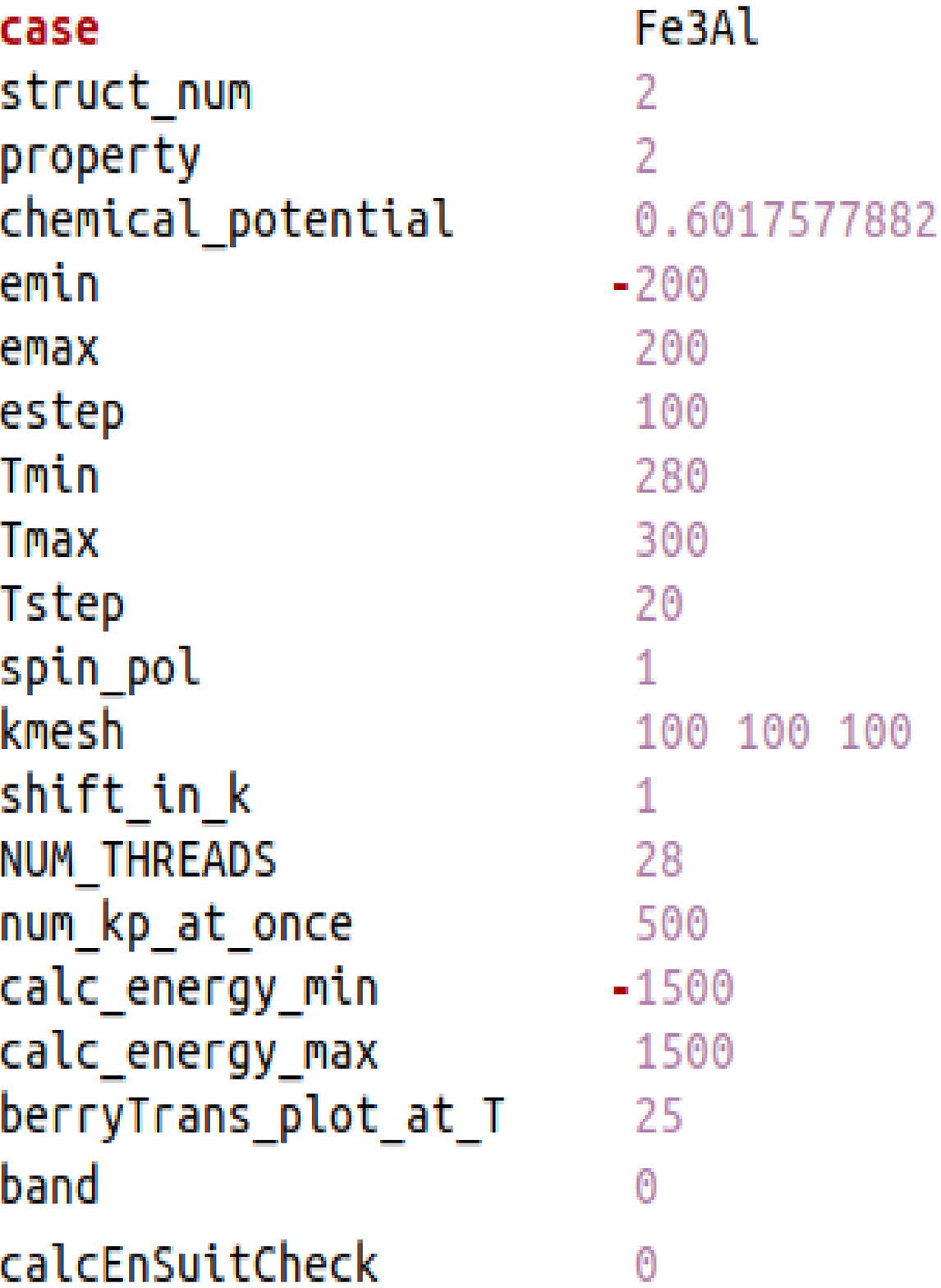}
        \caption{\textit{C-BerryTrans.input} file}
        \label{pmat}
    \end{subfigure}
    \begin{subfigure}[b]{0.55\textwidth}
        \includegraphics[width=1.00\textwidth,height=4.0cm]{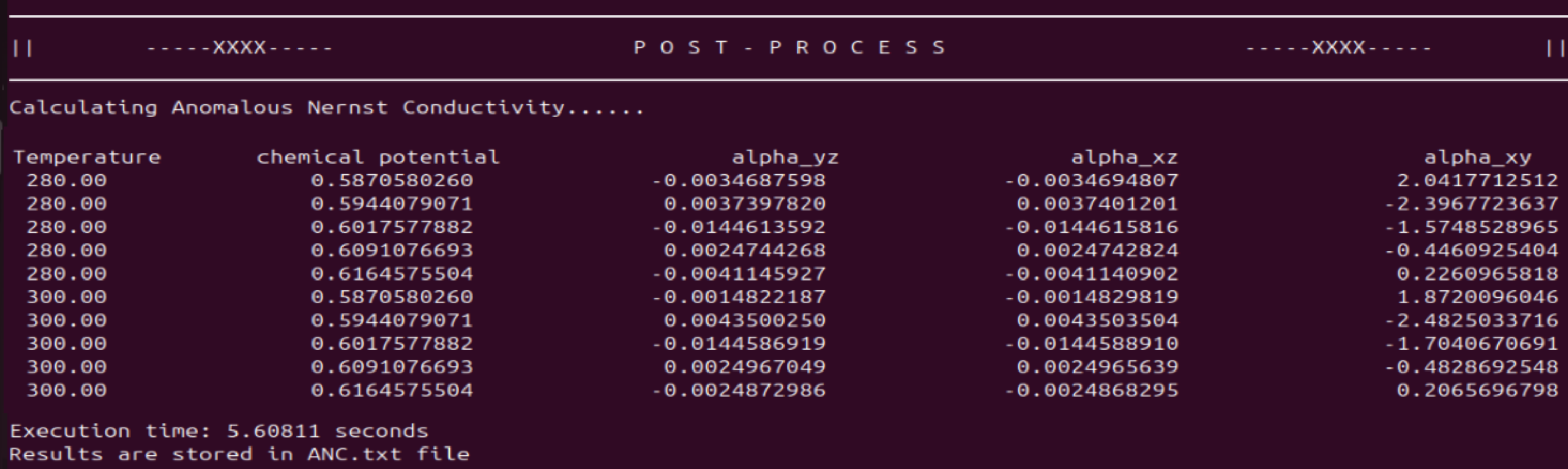}
        \caption{Output}
        \label{pmat}
    \end{subfigure}
    \caption{Figs. (a) and (b) shows that performing the post-process with larger size of temperature and chemical-potential window leads to \textit{free(): invalid size} errors. Figs. (c) and (d) depicts that the post-process is successfully carried out on reducing the size of temperature and chemical-potential window. Note: This is just a representative result. The post-process may still run on even larger temperature and chemical-potential window on the same computer. }
    \label{figFiles21}
\end{figure*}

\subsection{Post-process}

The post-process is accomplished by running the command: \textit{postBerryTrans}. It is important to note that when \textit{band=0}, the post-process involves summation over all \textbf{\textit{k}}-points of the full BZ to obtain the total AHC/ANC. In contrast, for \textit{band=1}, the code evaluates and outputs the $\boldsymbol\Omega$-weighted contributions along the specified \textbf{\textit{k}}-path without performing BZ integration. The prior case will be discussed now while the later one is discussed in section \ref{lab1}. The process starts with checking the input provided to the \textit{property} variable in \textit{C-BerryTrans.input} file. If the provided input is \textquoteleft 2', it will compute the value of $\Omega_{\mu \nu}^n(\textbf{\textit{k}}) \left[ (E_n - E_F) f_n(\textbf{\textit{k}}) + k_B T \ln \left(1 + \exp\left( \frac{E_n - E_F}{-k_B T} \right) \right) \right]$ for all the bands at each \textbf{\textit{k}}-points (for ANC). In a single run, it performs this calculations for a range of $T$ and $\omega$ values. For this, the $T$ range is decided by the values assigned to variables \textit{Tmin}, \textit{Tmax} and \textit{Tstep} while the $\omega$ window is given by the values provided to variables \textit{emin}, \textit{emax} and \textit{estep}. Finally, the module calculates $\alpha_{\mu \nu}^{\mathrm{ANC}}$ (for $\mu\nu$= {$xz$, $yz$ \& $xy$}) for each combination of $T$ and $\omega$ by summing up the respective values of $\Omega_{\mu\nu}^n(\textbf{\textit{k}}) \left[ (E_n - E_F) f_n(\textbf{\textit{k}}) + k_B T \ln \left(1 + \exp\left( \frac{E_n - E_F}{-k_B T} \right) \right) \right]$ for all the bands and across all the \textbf{\textit{k}}-points of the full BZ. The results obtained is mentioned in \textit{ANC.txt} file. Moving further, if the assigned value to the \textit{property} variable is \textquoteleft 1', the module will compute $f_n(\textbf{\textit{k}}) \Omega_{\mu \nu}^n(\textbf{\textit{k}})$ for all the bands at each \textbf{\textit{k}}-points (for AHC). After this, it will do the summation of this term corresponding to each value of $T$ and $\omega$ to obtain $\sigma_{\mu \nu}^{\mathrm{AHC}}$. The result will be stored in \textit{AHC.txt} file. It is important to note that while running the \textit{C-BerryTrans} code, the post-process is automatically carried out after the calculation of $\boldsymbol\Omega$. However, once the \textit{data.bin} files are created after the calculations of $\boldsymbol\Omega$, one has freedom to run post-process (by calling the command \textit{postBerryTrans}) for any range of $\omega$ and $T$ values by doing the respective modification in the \textit{C-BerryTrans.input} file. It is important to note that the post-processing should be performed using the same number of threads as those used during the $\boldsymbol\Omega$ calculations. Furthermore, total time taken for $\boldsymbol\Omega$ calculation is mentioned in \textit{final\_time.txt file}.\\\\
\textbf{Special Attention during Post-Processing:} During the post-processing stage (when \textit{band=0}), errors such as:
\begin{center}
\textit{munmap\_chunk(): invalid pointer} \\
\textit{Aborted (core dumped)}
\end{center}
or,
\begin{center}
\textit{free(): invalid size} \\
\textit{Aborted (core dumped)}
\end{center}
may occasionally appear. In some cases, minor inconsistencies may also be observed in the last few lines of the computed AHC/ANC values written in the \textit{AHC.txt} or \textit{ANC.txt} files. These issues do not indicate incorrect calculations. Instead, they generally arise from insufficient system resources (particularly RAM) when a very large number of temperature and chemical-potential combinations are processed within a single post-processing run.

\begin{figure*}[t]
    \centering
    \begin{subfigure}[b]{0.35\textwidth}
        \includegraphics[width=0.90\textwidth,height=2.5cm]{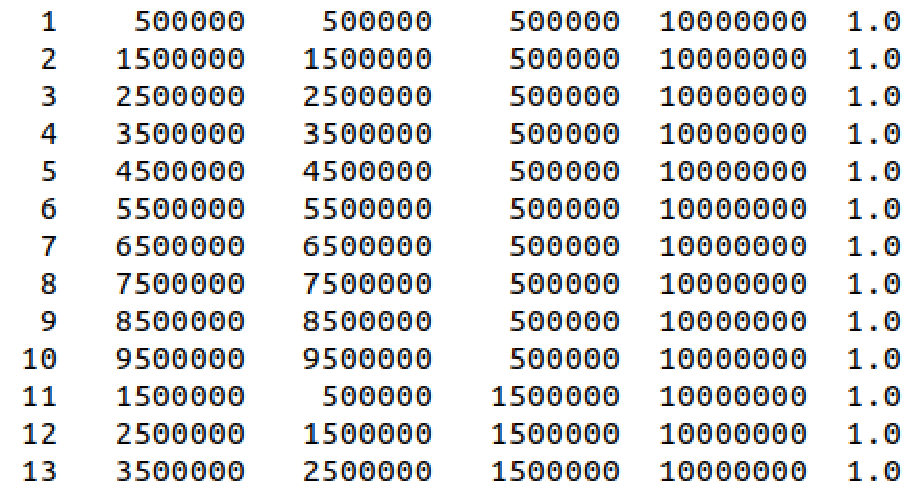}
        \caption{\textit{case.klist} file}
        \label{klist}
    \end{subfigure}
    \begin{subfigure}[b]{0.55\textwidth}
        \includegraphics[width=0.90\textwidth,height=2.5cm]{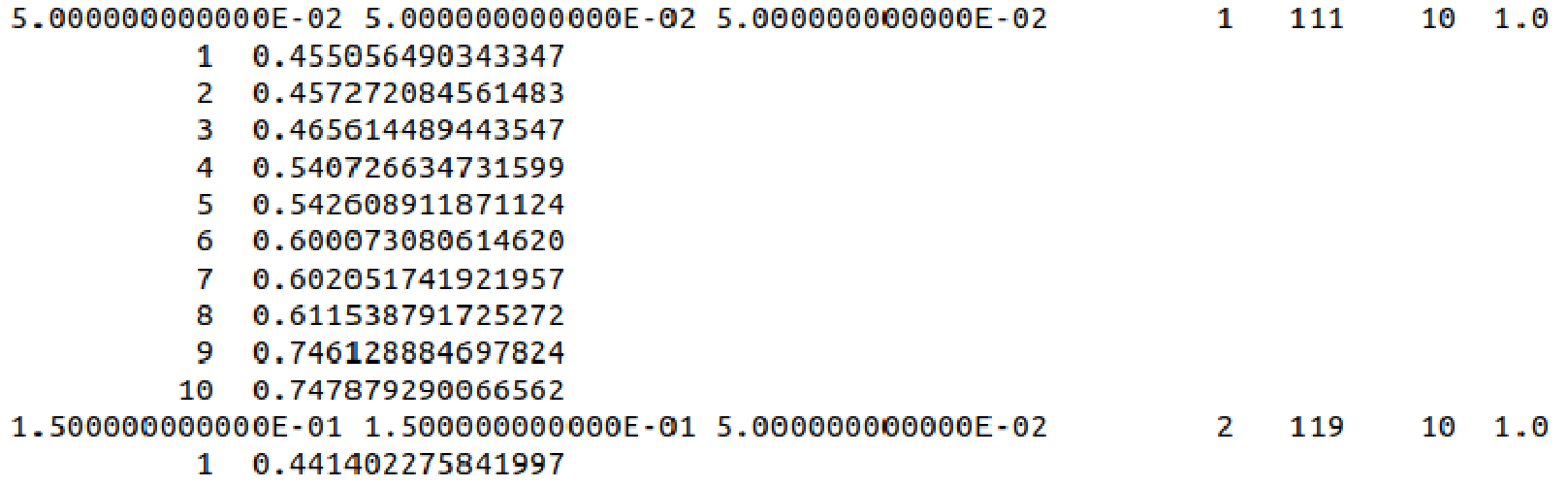}
        \caption{\textit{case.energy} file}
        \label{energy}
    \end{subfigure}
    \begin{subfigure}[b]{0.35\textwidth}
        \includegraphics[width=1.00\textwidth,height=2.5cm]{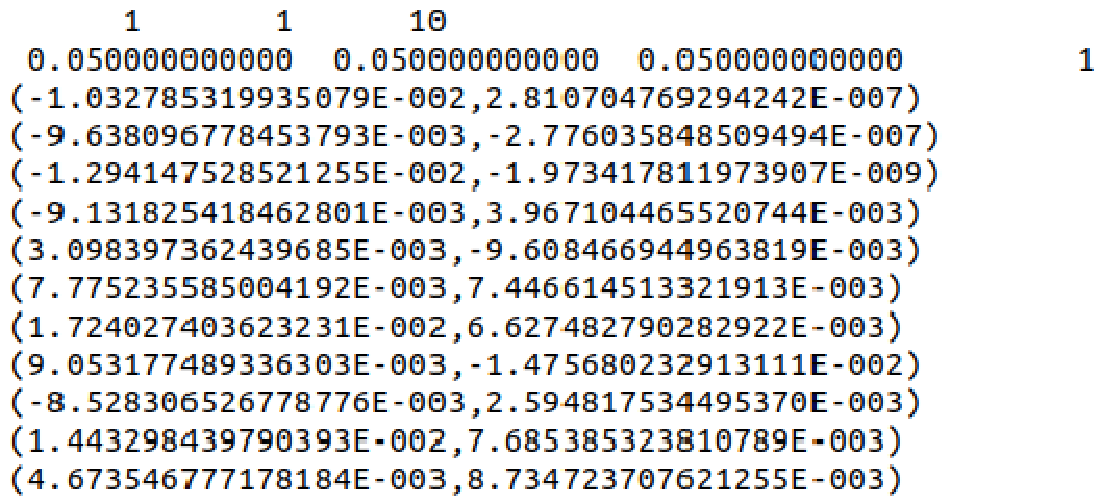}
        \caption{\textit{case.pmat} file}
        \label{pmat}
    \end{subfigure}
    \caption{The screenshots of the various files obtained from the WIEN2k calculations that are needed by the \textit{C-berryTrans} code for computing the $\boldsymbol\Omega$-driven transverse transport properties.}
    \label{figFiles2}
\end{figure*}

For example, consider the input parameters: \textit{emin} = $-200$, \textit{emax} = $200$, \textit{estep} = $10$, \textit{Tmin} = $0$, \textit{Tmax} = $300$, and \textit{Tstep} = $20$. The total number of temperature and chemical-potential combinations generated is
\[
\left(\frac{200 - (-200)}{10}\right) \times \left(\frac{300 - 0}{20}\right) = 40 \times 15 = 600
\]
Furthermore, each combination involves the evaluation of three tensor components of AHC/ANC. In such scenario, the memory requirement during post-processing can become significantly large. Consequently, systems with limited available RAM may encounter runtime interruptions or minor inconsistencies in the final few lines of the output files.

To avoid these problems, it is recommended to divide the temperature window and/or chemical-potential window into smaller segments and perform the post-processing calculations separately for each segment. The individual outputs can then be combined later to reconstruct the complete dataset. The error caused by using large temperature and chemical-potential windows in the case of Fe$_3$Al is illustrated in Fig.~\ref{figFiles21}.

It is important to note that the total post-processing computational time mainly depends on the overall number of temperature and chemical-potential combinations. Therefore, splitting the calculations into smaller segments does not increase the overall computational cost, but significantly improves stability and prevents memory-related errors. This approach also removes inconsistencies that may appear in the final lines of the output files. On computer systems with larger memory capacity, such issues may not arise.

\section{Technical details}

All the modules of \textit{C-BerryTrans} are written in C++, making use of standard libraries including \textit{cmath}, \textit{vector}, \textit{chrono}, \textit{omp.h}, \textit{cstdlib}, and \textit{string}. The calculation of $\boldsymbol\Omega$ (and AHC/ANC in the post-process) is parallelized over the \textbf{\textit{k}}-points using the OpenMP method. In addition, computation of $\boldsymbol\Omega$ and ANC is carried out using an \textit{ab-initio} approach. At present, the code is interfaced with the WIEN2k package, but it can be easily linked with other DFT packages such as ELK\cite{ELK}, Quantum Espresso\cite{giannozzi2009quantum}, etc. The implementation is fully general, without assumptions tied to any specific material class or model, thereby ensuring wide applicability across diverse systems.

To better understand the working of \textit{C-BerryTrans} code, it is important to discuss the format and indexing of three main WIEN2k files: the \textit{case.klist} file, the eigenvalue file (\textit{case.energyso}), and the momentum-matrix file (\textit{case.pmat}). These files serve as the core inputs for \textit{C-BerryTrans} calculations, and the following discussion clarifies how their contents are interconnected.

When the calculation is carried out with \textit{C-BerryTrans}, if \textit{spin\_pol} = 1, the code extracts the files \textit{case.energysoup} and \textit{case.pmatup} from the \textit{case} sub-directory of the respective \textit{process\_x} directory and renames them as \textit{case.energyso} and \textit{case.pmat}, respectively. It is important to note that when SOC is included, the files \textit{case.energysoup} and \textit{case.pmatup} contain information for both \textit{up} and \textit{down} spin.  If \textit{spin\_pol} = 0, the code directly takes out the \textit{case.energyso} and \textit{case.pmat} from the \textit{case} sub-directory without renaming. One should note that the format of \textit{case.energysoup} is identical to that of \textit{case.energyso}, and similarly, the format of \textit{case.pmatup} is identical to \textit{case.pmat}.

The format of \textit{case.klist} file is shown in Fig. \ref{klist}. Each line in this file corresponds to one \textbf{\textit{k}}-point. It is seen that each line contains 6 numeric entries. The first entry denotes the index of the \textbf{\textit{k}}-point. The following three integers, when divided by the fourth entry, yield the Cartesian coordinates of the \textbf{\textit{k}}-point. The last entry specifies the multiplicity of the \textbf{\textit{k}}-point in the BZ. Since \textit{C-BerryTrans} generates the \textbf{\textit{k}}-list over the full BZ, \textbf{\textit{k}}-point symmetry is not considered, and hence the multiplicity of all \textbf{\textit{k}}-points is set to 1. 

Moving next, the format of the \textit{case.energyso} file is shown in Fig. \ref{energy}. Each \textbf{\textit{k}}-point entry begins with a header line containing seven values. The first three values represent the cartesian coordinates of the \textbf{\textit{k}}-point. The fourth value denotes the \textbf{\textit{k}}-point index, consistent with the indexing in the \textit{case.klist} file. The fifth value is related to the diagonalization of the Hamiltonian at that \textbf{\textit{k}}-point (not relevant in the present context). The sixth value specifies the number of bands at the \textbf{\textit{k}}-point, and the last entry indicates the multiplicity. Following the header line, there are multiple lines, each containing the band index and the corresponding band energy at that \textbf{\textit{k}}-point. This pattern is repeated for every subsequent \textbf{\textit{k}}-point in the \textit{case.klist} file. Thus, the \textit{case.energyso} file provides the complete band-structure information at all \textbf{\textit{k}}-points listed in the \textit{case.klist} file.

\begin{table*}
\caption{\label{tab73}
{The various input details and the \textbf{\textit{k}}-mesh size used to calculate the ground state energy of the materials using WIEN2k package. The \textbf{\textit{k}}-meshes are taken in the irreducible part of the BZ (IBZ).}}
 \centering
 \small 
 \begin{tabular}{|c|c|c|c|c|c|}
   \hline
    {$\textbf{case}$} &{$\textbf{Space}$} &{$\textbf{Lattice}$} &{$\textbf{Wyckoff}$} &{\textbf{\textit{k}-mesh}} &{$\textbf{Exchange}$} \\
    {$\textbf{}$} &{$\textbf{Group}$} &{$\textbf{Parameters}$} &{$\textbf{Positions}$} &{(in IBZ)} &{$\textbf{Correlation}$} \\
   \hline
    
       Fe\cite{yao2004first}   &$Im\bar{3}m$   & a=b=c=2.87 \AA      & Fe = (0.00, 0.00, 0.00)             &  10$\times$10$\times$10 & PBESol\\
                     &          & $\alpha$=$\beta$=$\gamma$=90\degree       &           &                    &    \\ \hline
                     
      Fe$_3$Ge\cite{drijver1976magnetic,kanematsu1963magnetic,cao2009large} &$Fm\bar{3}m$ & a=b=c=5.76 \AA       & Fe (I) = (0.75,0.25,0.75)        & 10$\times$10$\times$10 & PBESol \\
         &           & $\alpha$=$\beta$=$\gamma$=90\degree & Fe (II) = (0.00, 0.00, 0.50)        &                 &        \\
         &           &                                    & Ge = (0.00, 0.00, 0.00)              &           &              \\ \hline

    Pd\cite{guo2014anomalous}   &$Fm\bar{3}m$   & a=b=c=3.89 \AA      & Pd = (0.00, 0.00, 0.00)             &  10$\times$10$\times$10 & PBE \\
                     &          & $\alpha$=$\beta$=$\gamma$=90\degree       &           &                     &  \\  \hline
Fe$_3$Al\cite{nishino1997semiconductorlike} &$Fm\bar{3}m$ & a=b=c=5.77 \AA       & Fe (I) = (0.00,0.00,0.50)        & 10$\times$10$\times$10 & PBESol \\
         &           & $\alpha$=$\beta$=$\gamma$=90\degree & Fe (II) = (0.75, 0.25, 0.75)        &               &          \\
         &           &                                    & Al = (0.00, 0.00, 0.00)              &           &              \\ \hline
Co$_2$FeAl\cite{shukla2022atomic} &$Fm\bar{3}m$ & a=b=c=5.70 \AA              & Co = (0.25, 0.25, 0.25)        & 10$\times$10$\times$10  & PBESol \\
         &           & $\alpha$=$\beta$=$\gamma$= 90\degree     & Fe = (0.50, 0.50, 0.50)        &         &                \\
         &           &                                   & Al = (0.00, 0.00, 0.00)              &         &                \\
    
   \hline
 \end{tabular}
 \label{tab73}
\end{table*}

\begin{table*}
\caption{\label{tab74}
{The details of the values assigned to the input parameters for the \textit{C-BerryTrans} code. The \textbf{\textit{k}}-mesh is sampled across the full BZ.}}
 \centering
 \small 
 \begin{tabular}{|c|c|c|c|c|c|c|}
   \hline
    {$\textbf{\textit{case}}$} &{$\textbf{\textit{struct\_num}}$} &{$\textbf{\textit{chemical\_potential}}$} &{$\textbf{\textit{spin\_pol}}$} &{$\textbf{\textit{kmesh}}$} &{$\textbf{\textit{shift\_in\_k}}$} &{$\textbf{\textit{property}}$} \\
    {} &{} &{$\textbf{(Rydberg)}$} &{} &{} &{} &{} \\
   \hline
       Fe     & 3 &  0.6211893080 & 1 & 400$\times$400$\times$400 & 1 & 1\\
   Fe$_3$Ge   & 2 &  0.6681367548 & 1 & 400$\times$400$\times$400 & 1 & 1\\
       Pd     & 2 &  0.5153375149 & 1 & 400$\times$400$\times$400 & 1 & 2\\
   Fe$_3$Al   & 2 &  0.6017577882 & 1 & 400$\times$400$\times$400 & 1 & 2\\
   Co$_2$FeAl & 2 &  0.6276267143 & 1 & 400$\times$400$\times$400 & 1 & 1 \& 2\\
   \hline
 \end{tabular}
 \label{tab74}
\end{table*}

The format of the \textit{case.pmat} file is shown in Fig. \ref{pmat}. It contains the momentum-matrix elements between the eigenstates at each \textbf{\textit{k}}-point. The first line of each \textbf{\textit{k}}-point block has three entries: the \textbf{\textit{k}}-point index, the starting band index, and the ending band index that fall within the energy window specified in the \textit{case.inop} file. In \textit{C-BerryTrans}, this energy window is automatically chosen to be consistent for both eigenenergy and momentum-matrix calculations. As a result, the band indices always begin from 1 and end at the total number of bands present at that \textbf{\textit{k}}-point, as listed in the \textit{case.energyso} file. Moving further, the second line in the file gives the cartesian coordinates of the \textbf{\textit{k}}-point along with its index in the \textit{case.klist} file. The subsequent lines list the momentum-matrix elements in the order $\langle n |p_x| m \rangle$, $\langle n |p_y| m \rangle$, and $\langle n |p_z| m \rangle$. Here, $| n \rangle$ and $| m \rangle$ indicate the Bloch states associated with band indices $n$ and $m$. Each momentum-matrix element is stored in the complex format $(a,b)$, where the first number $a$ represents the real part and the second number $b$ represents the imaginary part. The $(n,m)$ combinations start from the initial band index specified in the first line. For example, if the bands run from 1 to 10, the first three entries correspond to $\langle 1 |p_x| 1 \rangle$, $\langle 1 |p_y| 1 \rangle$, $\langle 1 |p_z| 1 \rangle$, followed by $\langle 1 |p_x| 2 \rangle$, $\langle 1 |p_y| 2 \rangle$, $\langle 1 |p_z| 2 \rangle$, and so on up to band 10. Importantly, the matrix elements $\langle n |p_x| m \rangle$ and $\langle m |p_x| n \rangle$ (similarly for $p_y$ and $p_z$) are related by complex conjugation. Therefore, only one of these pairs is present in the \textit{case.pmat} file. For instance, after $\langle 1 |p_x| 2 \rangle$, the file does not contain $\langle 2 |p_x| 1 \rangle$, since it is simply the complex conjugate. In this way, the number of momentum-matrix entries at a given \textbf{\textit{k}}-point is $3 \times \big(N + (N-1) + (N-2) + \cdots + 1\big)$, where $N$ is the number of bands at that \textbf{\textit{k}}-point. The same structure is then repeated for the next \textbf{\textit{k}}-point.

\subsection*{Checking convergence of $\Omega$-driven AHC/ANC}
For obtaining more realistic and physically sound result, it is important to check the convergence of $\boldsymbol\Omega$-driven AHC/ANC with respect to the number of unoccupied bands and the size of \textbf{\textit{k}}-mesh. In the default mode of WIEN2k, total number of bands are taken to be (\textit{number of electrons}*2 +5)/\textit{ispin}, where \textit{ispin} equals to 2 for non-spinpolarized case and 1 for spinpolarized case. This is expected to be enough and thus, for the validation of the code, we have not carried out the convergence test with respect to the number of unoccupied bands. Moreover, the results obtained were well-matching with the previously reported data (as shall be discussed later). However, one may check the convergence of $\boldsymbol\Omega$ and the related AHC/ANC with the change in the number of unoccupied bands, in case if the obtained result appears to be unphysical. In this regard, the number of bands to be included in the calculation can be modified in the \textit{case.in1} file. For more details, refer the WIEN2k userguide, section: \textit{LAPW1 (generates eigenvalues and eigenvectors)}\cite{WIEN2kUserManual}. After modifying the number of bands in \textit{case.in1} file, user needs to start the computation from the beginning. That is, perform the self-consistent ground state energy calculations and then carry-out the \textit{C-BerryTrans} computations.  Apart from the number of unoccupied bands, the user is strongly advised to check the convergence of AHC/ANC with respect to the size of \textbf{\textit{k}}-mesh. This can be done by computing the AHC/ANC with increasing values provided to the \textit{kmesh} variable in the \textit{C-BerryTrans.input} file and to check the changes in the obtained result. One must keep increasing the size of the \textbf{\textit{k}}-mesh till there is no more change in the values of AHC/ANC with further enhancement in the \textbf{\textit{k}}-mesh size.

\section{Test Cases}

\textit{C-BerryTrans} has been benchmarked over five ferromagnetic materials: Fe\cite{yao2004first}, Fe$_3$Ge\cite{drijver1976magnetic,kanematsu1963magnetic,cao2009large}, Pd\cite{guo2014anomalous}, Fe$_3$Al\cite{nishino1997semiconductorlike} and Co$_2$FeAl\cite{shukla2022atomic}. Fe is a body-centered cubic crystal while the other four materials crystallize in face-centered cubic crystal structure. These materials are expected to exhibit ANE and AHE due to their intrinsic magnetism. In this regards, previous theoretical and experimental works reports the $T$-dependent values of AHC in Fe\citep{yao2004first,dheer1967galvanomagnetic}, Fe$_3$Ge\cite{li2023anomalous} and Co$_2$FeAl\citep{huang2015anomalous}. Furthermore, previous works also report the ANC in Pd\citep{guo2014anomalous}, Fe$_3$Al\citep{sakai2020iron,koepernik2023symmetry} \& Co$_2$FeAl\citep{noky2020giant}. Thus, it is convincing to use them for benchmarking the \textit{C-BerryTrans} code.

\begin{figure*}[t]
    \centering
    \begin{subfigure}[b]{0.28\textwidth}
        \includegraphics[width=\textwidth]{fig3_1.eps}
        \caption{Pd}
        \label{alphaPd}
    \end{subfigure}
    \hspace*{0.02in}
    \begin{subfigure}[b]{0.28\textwidth}
        \includegraphics[width=\textwidth]{fig3_2.eps}
        \caption{Fe$_3$Al}
        \label{alphaFe3Al}
    \end{subfigure}
    \hspace*{0.02in}
    \begin{subfigure}[b]{0.28\textwidth}
        \includegraphics[width=\textwidth]{fig3_3.eps}
        \caption{Co$_2$FeAl}
        \label{alphaCo2FeAl}
    \end{subfigure}
    \caption{$\alpha^{ANC}_{xy}$ as a function of $\omega$ for the respective materials at 300 $K$. The value of $\omega$ is scaled with respect to Fermi energy.}
    \label{figFiles2}
\end{figure*}
 
\subsection{Computational Details}
The self-consistent ground state energy calculations for all materials are carried out using the WIEN2k package. SOC is included in these calculations. Since all the materials exhibit magnetic behavior, spin-polarization is also taken into account, with the magnetization aligned along the \textit{z}-axis. Details such as the space group, lattice parameters, atomic Wyckoff positions, \textbf{\textit{k}}-mesh size, and the exchange–correlation functional used in the ground state calculations are provided in table \ref{tab73}. BZ is sampled over its irreducible wedge in all the ground state calculations. The energy convergence criterion for the self-consistent field iterations is set to 10$^{-4}$ Rydberg per unit cell.

For the calculations of intrinsic transverse response, the \textbf{\textit{k}}-mesh size and values assigned to various input parameters of the \textit{C-BerryTrans} code are listed in table \ref{tab74}. The $\omega$-dependent values of $\alpha_{xy}^{ANC}$/$\sigma_{xy}^{AHC}$ are calculated within energy range of -200 to 200 meV relative to the Fermi level. The $\omega$ window is divided into 40 parts. Accordingly, the parameters \textit{emin}, \textit{emax}, \textit{estep}, \textit{Tmin}, \textit{Tmax} and \textit{Tstep} are set to -200, 200, 10, 300, 300 and 25 (any non-zero value), respectively. Furthermore, for $T$-dependent calculations (ANC: within 25-300 $K$ and AHC: within 0-300 $K$) at the Fermi energy, the parameters \textit{emin}, \textit{emax}, \textit{estep}, \textit{Tmax} and \textit{Tstep} are set to 0, 0, 1, 300 and 25, respectively. For this computation, \textit{Tmin} is taken to be 0 for AHC and 25 for ANC calculations.  All computations are performed on a system with 32 GB RAM, using \textit{NUM\_THREADS} = 28 and \textit{num\_kp\_at\_once} = 5000. For these computation, assign value to the \textit{band} parameter as 0.

\begin{table}
\caption{\label{tab75}
{Convergence of $\alpha_{xy}^{ANC}$ in Fe$_3$Al and $\sigma_{xy}^{AHC}$ in Fe at 300 $K$ with respect to the size of \textbf{\textit{k}}-mesh. The value of $\omega$ is set at Fermi energy.}} 
 \centering
 \small 
 \begin{tabular}{|c|c|c|}
   \hline
    {\textbf{\textit{k}-mesh}} &{\textbf{$\alpha_{xy}^{ANC}$ for Fe$_3$Al}} &{\textbf{ $\sigma_{xy}^{AHC}$ for Fe}} \\
    {\textbf{}} &{\textbf{$AK^{-1}m^{-1}$}} &{\textbf{$S/cm$}} \\
   \hline
 50$\times$50$\times$50              & -2.12 & 831.96\\
 100$\times$100$\times$100           & -1.80 & 756.88\\
 150$\times$150$\times$150           & -1.87 & 739.12\\
 200$\times$200$\times$200           & -1.84 & 744.95\\
 250$\times$250$\times$250           & -1.80 & 740.41\\
 300$\times$300$\times$300           & -1.82 & 743.75\\
 350$\times$350$\times$350           & -1.85 & 743.46  \\
 400$\times$400$\times$400           & -1.85 & 743.49    \\
\hline
 \end{tabular}
 \label{tab75}
\end{table}

\subsection{Results and discussion}
At the very first, the \textbf{\textit{k}}-point convergence of $\alpha_{xy}^{ANC}$ has been carried out for Fe$_3$Al. For this, the value of $\omega$ is set at the Fermi energy while the $T$ is taken to be 300 $K$. The obtained result is presented in table \ref{tab75}. With increase in the size of \textbf{\textit{k}}-mesh, a significant variation in the value of $\alpha_{xy}^{ANC}$ is observed when the size of \textbf{\textit{k}}-meshes were small. However, beyond 350$\times$350$\times$350 \textbf{\textit{k}}-mesh size, the value of $\alpha_{xy}^{ANC}$ gets converged to -1.85 $AK^{-1}m^{-1}$. In addition to this, the \textbf{\textit{k}}-mesh convergence of $\sigma_{xy}^{AHC}$ is carried out for Fe. The values of $\omega$ and $T$ were set at the Fermi level and 300 $K$, respectively. Obtained results are depicted in table \ref{tab75}. It is seen that values of $\sigma_{xy}^{AHC}$ for Fe also get converged beyond the \textbf{\textit{k}}-mesh size of 350$\times$350$\times$350. Thus, these calculations provide an estimate of the size of \textbf{\textit{k}}-mesh required for ANC or AHC calculations using the present code. To ensure the accuracy of the results, all further computations of these quantities are performed on 400$\times$400$\times$400 size of \textbf{\textit{k}}-mesh.

The code is firstly benchmarked over Palladium (Pd). Calculations of $\omega$-dependent $\alpha_{xy}^{ANC}$ is carried out in the energy range of -200 to 200 meV around the Fermi energy. The $T$ value is set to 300 $K$ for these computations. Obtained results are shown in Fig. \ref{alphaPd}. The Fermi energy is scaled to 0 meV. It is seen that within the given $\omega$ range, $\alpha_{xy}^{ANC}$ is minimum ($\sim$-1.85 $AK^{-1}m^{-1}$) when the $\omega$ value is 200 meV below the Fermi level. With the rise in $\omega$, $\alpha_{xy}^{ANC}$ increases and attains a maximum value of $\sim$1.29 $AK^{-1}m^{-1}$ at $\omega$ equals to $\sim$40 meV above the Fermi energy. On further rise in value of $\omega$, $\alpha_{xy}^{ANC}$ is found to increase. In addition to this, the $T$-dependent values of $\alpha_{xy}^{ANC}$ are also computed within 25-300 $K$, when the $\omega$ value corresponds to Fermi energy. Obtained result is shown in table \ref{tabb4}. It is found that with the rise in the value of $T$, the magnitude of $\alpha_{xy}^{ANC}$ increases. At 300 $K$, the value of $\alpha_{xy}^{ANC}$ is obtained to be 0.97 $AK^{-1}m^{-1}$. Guo \textit{et. al.}, have previously studied ANC in Pd using computational approach\cite{guo2014anomalous}. In their work, the reported magnitude of $\alpha_{xy}^{ANC}$ at 300 $K$ is 0.72 $AK^{-1}m^{-1}$. Thus, the obtained result from \textit{C-BerryTrans} code is found to be in a very good match with the reported data.

\begin{table}
\caption{\label{tab:table1}%
\normalsize{Value of $\alpha_{xy}^{ANC}$ vs $T$. The value of $\omega$ is set at Fermi level.
}}
\begin{tabular}{|c|c|c|c|}
\hline
\textrm{\textbf{$T$}}&
\textrm{\textbf{}}&
\textrm{\textbf{$\alpha_{xy}^{ANC}$}}&\\
\colrule
\textrm{\textbf{}}&
\textrm{\textbf{Pd}}&
\textrm{\textbf{Fe$_3$Al}}&
\textrm{\textbf{Co$_2$FeAl}}\\
\textrm{\textbf{($K$)}}&
\textrm{\textbf{($AK^{-1}m^{-1}$})}&
\textrm{\textbf{($AK^{-1}m^{-1}$})}&
\textrm{\textbf{($AK^{-1}m^{-1}$})}\\
\colrule
 25           &  0.07  & -0.05 & 0.01 \\
 50           &  0.14  & -0.14 & 0.02 \\
 75           &  0.23  & -0.24 & 0.03  \\
 100          &  0.32  & -0.37 & 0.04   \\
 125          &  0.42  & -0.53 & 0.04  \\
 150          &  0.52  & -0.71 & 0.05    \\
 175          &  0.62  & -0.91 & 0.06   \\
 200          &  0.71  & -1.11 & 0.07  \\
 225          &  0.79  & -1.31 & 0.08  \\
 250          &  0.86  & -1.50 & 0.09  \\
 275          &  0.92  & -1.68 & 0.10 \\
 300          &  0.97  & -1.85 & 0.10   \\
\hline
\end{tabular}
\label{tabb4}
\end{table}

Nextly, using the \textit{C-BerryTrans} code, the room temperature value of $\alpha_{xy}^{ANC}$ is calculated for Fe$_3$Al corresponding to various values of $\omega$ ranging from -200 to 200 meV around the Fermi level. Obtained results are shown in Fig. \ref{alphaFe3Al}. Within the given $\omega$ window, the value of $\alpha_{xy}^{ANC}$ is found to first decrease to attain a minimum value of -2.84 $AK^{-1}m^{-1}$ (at $\omega$ corresponding to 60 meV below the Fermi energy). With further rise in $\omega$, value of $\alpha_{xy}^{ANC}$ is seen to increase. In addition to this, the $T$-dependent values of $\alpha_{xy}^{ANC}$ is computed for the material. The obtained results are shown in table \ref{tabb4}. It is found that, within 25-300 $K$, magnitude of $\alpha_{xy}^{ANC}$ rises with the increase in $T$. At 300 $K$, the value is found to be -1.85 $AK^{-1}m^{-1}$. Sakai \textit{et. al.} have previously computed the $T$-dependent value of $\alpha_{xy}^{ANC}$ using high-throughput computation\cite{sakai2020iron}. In their work, the maximum magnitude of $\alpha_{xy}^{ANC}$ for $T$ values ranging up to 500 $K$ is reported to be 2.7 $AK^{-1}m^{-1}$. For the same $T$ range (\textit{i.e.}, $T$ $\leq$ 500 $K$), the maximum magnitude of $\alpha_{xy}^{ANC}$ was reported to be 3.0 $AK^{-1}m^{-1}$ in another work\cite{koepernik2023symmetry}. In this $T$ range, using \textit{C-BerryTrans} code, the maximum magnitude of $\alpha_{xy}^{ANC}$ is found to be  2.83 $AK^{-1}m^{-1}$ (when $\omega$ is set at the Fermi energy). It is also important to highlight here that the calculations in the above mentioned works were done using tight-binding models obtained from the wannierization method. Considering this aspect, the result obtained from \textit{C-BerryTrans} code is found to be in a good match with the reported data.

\begin{figure*}[t]
    \centering
    \begin{subfigure}[b]{0.27\textwidth}
        \includegraphics[width=\textwidth]{fig4_1.eps}
        \caption{Fe}
        \label{sigmaFe}
    \end{subfigure}
    \hspace*{0.065in}
    \begin{subfigure}[b]{0.27\textwidth}
        \includegraphics[width=\textwidth]{fig4_2.eps}
        \caption{Fe$_3$Ge}
        \label{sigmafe3ge}
    \end{subfigure}
    \hspace*{0.068in}
    \begin{subfigure}[b]{0.27\textwidth}
        \includegraphics[width=\textwidth]{fig4_3.eps}
        \caption{Co$_2$FeAl}
        \label{figco2feal}
    \end{subfigure}
    \caption{The black (red) curve represents the $\sigma_{xy}^{AHC}$ vs $\omega$ for respective materials at 0 (300) $K$. The $\omega$ is scaled with respect to Fermi energy.}
    \label{figFiles3}
\end{figure*}

\begin{figure}[t]
    \centering
    \includegraphics[width=0.40\textwidth,height=4.3cm]{fig4.eps}
    \caption{$\sigma_{xy}^{AHC}$ vs $\omega$ for Co$_2$FeAl at 2 $K$. The $\omega$ is scaled with respect to Fermi energy. }
    \label{sigmaCo2FeAl}
\end{figure}

Moving next, the \textit{C-BerryTrans} code is tested over Co$_2$FeAl. As discussed above, $\omega$-dependent $\alpha_{xy}^{ANC}$ were computed within -200 to 200 meV around the Fermi energy and the obtained results are shown in Fig. \ref{alphaCo2FeAl}. Value of $T$ is taken to be 300 $K$. It is seen that as one traces the $\omega$ values from -200 to 200 meV around the Fermi energy, value of $\alpha_{xy}^{ANC}$ first increases to attain a maxima ($\sim$0.10 $AK^{-1}m^{-1}$) at the Fermi energy. With further rise in the value of $\omega$, $\alpha_{xy}^{ANC}$ again starts to decrease to attain a minima ($\sim$0.00 $AK^{-1}m^{-1}$) at $\omega$ equals to $\sim$180 meV above the Fermi energy. Beyond this, the value of $\alpha_{xy}^{ANC}$ starts to increase again with the rise in $\omega$. In addition to this, the computed values of $T$-dependent $\alpha_{xy}^{ANC}$ within the $T$ range of 25-300 $K$ is mentioned in table \ref{tabb4}. The value of $\alpha_{xy}^{ANC}$ is found to increase with the rise in $T$. The value is found to be $\sim$0.10 $AK^{-1}m^{-1}$ at 300 $K$. Noky \textit{et. al.} have previously calculated $\alpha_{xy}^{ANC}$ in Co$_2$FeAl using high-throughput calculations (involving wannierization techniques) and found its magnitude to be 0.06 $AK^{-1}m^{-1}$ at 300 $K$\cite{noky2020giant}. Apart from ANC, the $T$-dependent variation of magnitude of $\sigma_{xy}^{AHC}$ is also studied using the \textit{C-BerryTrans} code. The obtained results are mentioned in table \ref{tabb44}. It is found that magnitude of $\sigma_{xy}^{AHC}$ is nearly constant with the change in the values of $T$ within 0-300 $K$. In addition to this, the effect of change in the value of $\omega$ on $\sigma_{xy}^{AHC}$ of Co$_2$FeAl is studied for $T$ values of 0 $K$ \& 300 $K$. Obtained results are depicted in Fig. \ref{figco2feal}. It is observed that, within the given $\omega$-window, $\sigma_{xy}^{AHC}$ at 0 $K$ \& 300 $K$ are very close to each other. Moreover, $\sigma_{xy}^{AHC}$ show slow variation with the change in $\omega$ within the given $\omega$-window. In the literature survey, we came across a work in which $\sigma_{xy}^{AHC}$ is calculated at 2 $K$ (42 $S/cm$\cite{shukla2022atomic}). Thus, calculation of $\sigma_{xy}^{AHC}$ is carried out at 2 $K$ using \textit{C-BerryTrans} code. Obtained result is shown in Fig. \ref{sigmaCo2FeAl}. At the Fermi level, the magnitude of $\sigma_{xy}^{AHC}$ is found to be $\sim$56 $S/cm$ at 2 $K$. The features and magnitudes of the results are in good agreement with the previously reported theoretical result (39 $S/cm$\cite{huang2015anomalous}) at 2 $K$. Thus, the calculated value of $\alpha_{xy}^{ANC}$ and $\sigma_{xy}^{AHC}$ from \textit{C-BerryTrans} code is found to be in fairly good match with the reported data.

\begin{table}
\caption{\label{tab:table1}%
\normalsize{Magnitude of $\sigma_{xy}^{AHC}$ vs $T$. The value of $\omega$ is set at Fermi level.
}}
\begin{tabular}{|c|c|c|c|}
\hline
\textrm{\textbf{$T$}}&
\textrm{\textbf{}}&
\textrm{\textbf{$\sigma_{xy}^{AHC}$}}&\\
\colrule
\textrm{\textbf{}}&
\textrm{\textbf{Fe}}&
\textrm{\textbf{Fe$_3$Ge}}&
\textrm{\textbf{Co$_2$FeAl}}\\
\textrm{\textbf{($K$)}}&
\textrm{\textbf{($S/cm$)}}&
\textrm{\textbf{($S/cm$)}}&
\textrm{\textbf{($S/cm$)}}\\
\colrule
 0            & 775.94  & 322.90 & 56.15 \\
 25           & 774.61  & 323.10 & 56.29 \\
 50           & 773.33  & 324.44 & 56.48 \\
 75           & 772.17  & 326.08 & 56.49  \\
 100          & 770.78  & 326.99 & 56.49   \\
 125          & 768.97  & 326.80 & 56.49  \\
 150          & 766.62  & 325.63 & 56.51    \\
 175          & 763.72  & 323.74 & 56.52   \\
 200          & 760.35  & 321.42 & 56.55  \\
 225          & 756.59  & 318.87 & 56.57  \\
 250          & 752.55  & 316.24 & 56.60  \\
 275          & 748.32  & 313.60 & 56.63 \\
 300          & 744.00  & 311.00 & 56.66   \\
\hline
\end{tabular}
\label{tabb44}
\end{table}

Moving further, computation of the $T$-dependent value of $\sigma_{xy}^{AHC}$ at the Fermi level has been also carried out for Fe. The $T$ range chosen is from 0-300 $K$. The obtained results are mentioned in table \ref{tabb44}. It is seen from the table that value of $\sigma_{xy}^{AHC}$ decreases with the rise in the value of $T$. In addition to this, the calculations have been also carried out to study the variation of $\sigma_{xy}^{AHC}$ with the change in $\omega$. The values have been calculated for a range of $\omega$ values ranging from -200 meV to 200 meV with respect to Fermi energy. Furthermore, this calculation has been performed at 0 \& 300 $K$. Obtained results are shown in Fig. \ref{sigmaFe}. It is observed that, corresponding to both the the values of $T$, as $\omega$ value is raised above (or lowered below) the Fermi energy, magnitude of $\sigma_{xy}^{AHC}$ decreases. At the Fermi energy, magnitude of $\sigma_{xy}^{AHC}$ is obtained to be $\sim$775 ($\sim$744) $S/cm$ at 0 (300) $K$. Yao \textit{et. al.} have previously calculated the value of $\sigma_{xy}^{AHC}$ using the \textit{first-principles} approach combined with the adaptive mesh refinement method\cite{yao2004first}. The value of $\sigma_{xy}^{AHC}$ reported in their work is 751 (734) $S/cm$ at 0 (300) $K$. The value of $\sigma_{xy}^{AHC}$ calculated from \textit{C-BerryTrans} code is found to be in good agreement with the reported data.

Lastly the code is tested over Fe$_3$Ge. The $T$-dependent value of $\sigma_{xy}^{AHC}$ for $\omega$ corresponding to Fermi level is computed using \textit{C-BerryTrans} code. Obtained results are depicted in table \ref{tabb44}. It is seen from the table that with the rise in $T$ value from 0 to 100 $K$, magnitude of $\sigma_{xy}^{AHC}$ slowly increases from $\sim$322 $S/cm$ to $\sim$326 $S/cm$. With further increase in $T$, magnitude of $\sigma_{xy}^{AHC}$ falls down slowly. At 300 $K$, the computed value of $\sigma_{xy}^{AHC}$ is found to be 311 $S/cm$. Li \textit{et. al.} have explored AHC of Fe$_3$Ge using theoretical approach based on wannierization techniques\cite{li2023anomalous}. The value of AHC was calculated to be 227 \textit{S/cm} at 300 \textit{K}. Keeping in mind that they have used the wannierization techniques, the obtained value of $\sigma_{xy}^{AHC}$ from \textit{C-BerryTrans} code is in good match with their reported data. In addition to this, the effect of change in $\omega$ value on $\sigma_{xy}^{AHC}$ have been also explored for the material at 0 $K$ \& 300 $K$ using the \textit{C-BerryTrans} code. Obtained results are shown in Fig. \ref{sigmafe3ge}. It is seen that, within the given range of $\omega$ and at 0 $K$ \& 300 $K$, on raising the value of $\omega$ from -200 meV (with respect to Fermi energy), magnitude of $\sigma_{xy}^{AHC}$ first increase till $\omega$ value reaches close to -100 meV (with respect to Fermi energy). However, with further rise in the value of $\omega$, magnitude of $\sigma_{xy}^{AHC}$ falls down.


It is necessary to highlight here that the calculated values of ANC or AHC from \textit{C-BerryTrans} code may sometime deviate largely from the available experimental results. The possible reason for this may be the presence of defect or impurity in the sample. It may also arise if the sample is in polycrystalline phase. These conditions lead to various scattering mechanisms which is expected to be the cause for such deviations. Thus, user is suggested to consider these aspects while comparing the calculated results with the available experimental data.

\section{$\Omega$-driven AHC/ANC contribution along \textbf{\textit{k}}-path} 
\label{lab1}

While the AHC and ANC are obtained from BZ integrations of $\boldsymbol\Omega$, it is often highly informative to analyze the \textbf{\textit{k}}-resolved contribution entering these transport coefficients along high-symmetry paths\cite{yao2004first}. In the present implementation, the plotted quantity is not the bare $\boldsymbol\Omega(\textbf{\textit{k}})$, but the weighted contribution that directly enters the transport integrals.

For the AHC, the relevant quantity is $\sum_n f_n(\textbf{\textit{k}}) \Omega_{\mu \nu}^n(\textbf{\textit{k}})$ while for ANC, the corresponding contribution is $\sum_n \Omega_{\mu \nu}^n(\textbf{\textit{k}}) \left[ (E_n - E_F) f_n(\textbf{\textit{k}}) + k_B T \ln \left(1 + \exp\left( \frac{E_n - E_F}{-k_B T} \right) \right) \right]$. Resolving these weighted quantities along high-symmetry \textbf{\textit{k}}-paths enables a direct correlation between band dispersion and the dominant microscopic sources of the transverse transport response. Sharp features typically originate from avoided crossings, near-degenerate bands, or SOC-induced splittings close to the Fermi level\cite{xiao2010berry,helman2021anomalous,wang2006ab}.

To facilitate this analysis, \textit{C-BerryTrans} includes two dedicated modules (\textit{OmegaKpathAHC.cpp} \& \textit{OmegaKpathANC.cpp}) that computes these $\boldsymbol\Omega$-weighted contributions along user-defined high-symmetry paths. The workflow for activating this feature is described below.

\subsection*{Generation of \textit{case.klist\_band} using XCrySDen}

The \textbf{\textit{k}}-path file \textit{case.klist\_band} can be conveniently generated using the open-source visualization software \textsc{XCrySDen}\cite{KOKALJ1999176}. This approach provides an intuitive graphical interface for selecting high-symmetry paths in the BZ and is particularly useful for band-structure and $\boldsymbol\Omega$ analysis.

First, \textsc{XCrySDen} can be installed using:
\begin{center}
\textit{sudo apt-get install xcrysden}
\end{center}
To generate the \textit{case.klist\_band} file, the following steps should be followed:
\begin{enumerate}
    \item In the working directory, ensure that the standard structure file \textit{case.struct }(obtained from WIEN2k DFT calculations) is present.
    \item Execute the command:   \textit{xcrysden --wien\_kpath case.struct}.
    \item A graphical window displaying the BZ will appear, providing options to select high-symmetry \textbf{\textit{k}}-point paths.
    \item Choose the desired high-symmetry path segments and click the \textit{OK} button.
    \item A new dialog window will prompt for the total number of \textbf{\textit{k}}-points along the selected path. Enter the required number and click \textit{OK}.
    \item Subsequently, a file-save dialog will appear asking for the output filename. The file \emph{must} be saved with the exact name:  \textit{case.klist\_band}. For instance if the \textit{case} directory name if \textit{Fe}, then name must be \textit{Fe.klist\_band}. No other filename should be used.
    \item Click \textit{OK} to finalize the generation of the \textit{case.klist\_band} file.
\end{enumerate}

Once generated, the \textit{case.klist\_band} file should be kept in the same working directory where \textit{C-BerryTrans} is executed. This file is then read automatically when the input flag \textit{band = 1} is enabled in the \textit{C-BerryTrans.input} file.

\subsection*{Activation of the \textbf{\textit{k}}-path module}

The same workflow shown in Fig. \ref{workflow} also depicts the procedure to compute the $\boldsymbol{\Omega}$-weighted AHC/ANC contribution along a specified \textbf{\textit{k}}-path. As discussed before, assigning \textit{band = 1} activates the \textbf{\textit{k}}-path mode, where the code evaluates the $\boldsymbol{\Omega}$-weighted AHC/ANC contribution along the user-defined \textbf{\textit{k}}-path specified in the \textit{case.klist\_band} file. Furthermore, if spin-polarized calculation is required, the \textit{spin\_pol} variable must be set to 1; otherwise, it should be set to 0. For the \textbf{\textit{k}}-path calculation, the modified \textit{C-BerryTrans.input} file, the \textit{case.klist\_band} file, and the \textit{case} directory containing the self-consistently converged WIEN2k files must be placed inside the \textit{working} directory. After this, the calculation is executed using: \textit{berryTrans}. In this mode, the program performs:
\begin{itemize}
    \item SOC included band structure calculation along the points listed in \textit{case.klist\_band}. This is done in the \textit{bandCalc} directory. In the resulting band structure plot (\textit{case.bands.agr}), the Fermi energy is scaled with respect to the value assigned to the input variable \textit{chemical\_potential}.
    \item Evaluation of $\boldsymbol\Omega$ components $\Omega_{xy}$, $\Omega_{yz}$, and $\Omega_{xz}$ at each \textbf{\textit{k}}-point of the \textit{case.klist\_band}. This is done in the \textit{OMEGA\_calc\_dir}. The process is made parallel in a similar fashion as in the case of AHC/ANC computation. After the $\boldsymbol\Omega$ calculations, the \textit{postBerryTrans} module is automatically called. This computes the components (\textit{xy}, \textit{yz} \& \textit{xz}) of $\boldsymbol\Omega$-derived AHC/ANC along the \textbf{\textit{k}}-path defined in \textit{case.klist\_band} file. If \textit{property}=1, AHC computation is done else if \textit{property}=2, ANC is calculated. The temperature and chemical potential for the calculation is decided by the input provided to the variables \textit{Tmin} and \textit{chemical\_potential}, respectively. Final result is stored in the \textit{OmegaVsKpath.txt} file. It is important to note that once the band-resolved $\boldsymbol\Omega$ data are generated and stored, the post-process can be repeated for any desired combination of temperature and chemical-potential values by appropriately modifying the values of input parameters (\textit{i.e.}, \textit{Tmin} \& \textit{chemical\_potential}) and executing the \textit{postBerryTrans} command again.
\end{itemize}

\subsection*{Output files}

Upon successful execution, the following files are generated:

\begin{itemize}
    \item \textit{case.bands.agr}: Band structure data along the selected \textbf{\textit{k}}-path.
    \item \textit{OmegaVsKpath.txt}: $\boldsymbol\Omega$-derived AHC/ANC components at each \textbf{\textit{k}}-point along the path. The file has eight columns- chemical potential, Cartesian coordiantes of \textbf{\textit{k}}-points, \textit{length} of each \textbf{\textit{k}}-points with respect to first \textbf{\textit{k}}-point in the \textit{case.klist\_band} file, components (\textit{xy}, \textit{yz} \& \textit{xz}) of $\boldsymbol\Omega$-derived AHC/ANC. The data $\boldsymbol\Omega$-derived AHC/ANC vs \textit{length} can be directly plotted together with the band structure to identify $\boldsymbol\Omega$ \textquotedblleft hot spots”.
\end{itemize}

\subsection*{Example: Ferromagnetic bcc Fe}

As a representative example, we consider ferromagnetic bcc Fe. $\boldsymbol\Omega$-derived AHC calculation is carried out at 300 \textit{K} and 0.6211893080 chemical potential value. Fig. \ref{fig7} shows the calculated band structure and the corresponding $\Omega_{xy}^{AHC}(\textbf{\textit{k}})$ ($\Omega_{xy}(\textbf{\textit{k}})$-driven AHC) along the chosen high-symmetry path. Pronounced peaks in $\Omega_{xy}^{AHC}$ are observed near avoided band crossings and near-degenerate regions close to the Fermi energy. These features originate from enhanced interband matrix elements and small energy denominators in the Kubo formula expression for $\Omega_{xy}^{AHC}(\textbf{\textit{k}})$.

\begin{figure}[t]
    \centering
    \includegraphics[width=0.40\textwidth,height=5.0cm]{bands.eps}
    \caption{Band structure of Fe near the Fermi energy (upper panel) and the corresponding $\Omega_{xy}^{AHC}(\textbf{\textit{k}})$ (lower panel) along high-symmetry lines.}
    \label{fig7}
\end{figure}

In Fe, the dominant contribution to $\Omega_{xy}^{AHC}(\textbf{\textit{k}})$ originates due to SOC splitted bands with one beign occupied and the being unoccupied. The magnitude and peak position obtained using the \textit{C-BerryTrans} code is in good agreement with the previous reported data\cite{yao2004first}). This result highlights that the module can serve as an important diagnostic tool for identifying $\boldsymbol\Omega$ hot spots and understanding the microscopic origin of $\boldsymbol\Omega$-driven transverse transport in magnetic and topological materials.

\begin{figure*}[t]
    \centering
    \includegraphics[width=0.85\textwidth,height=5.0cm]{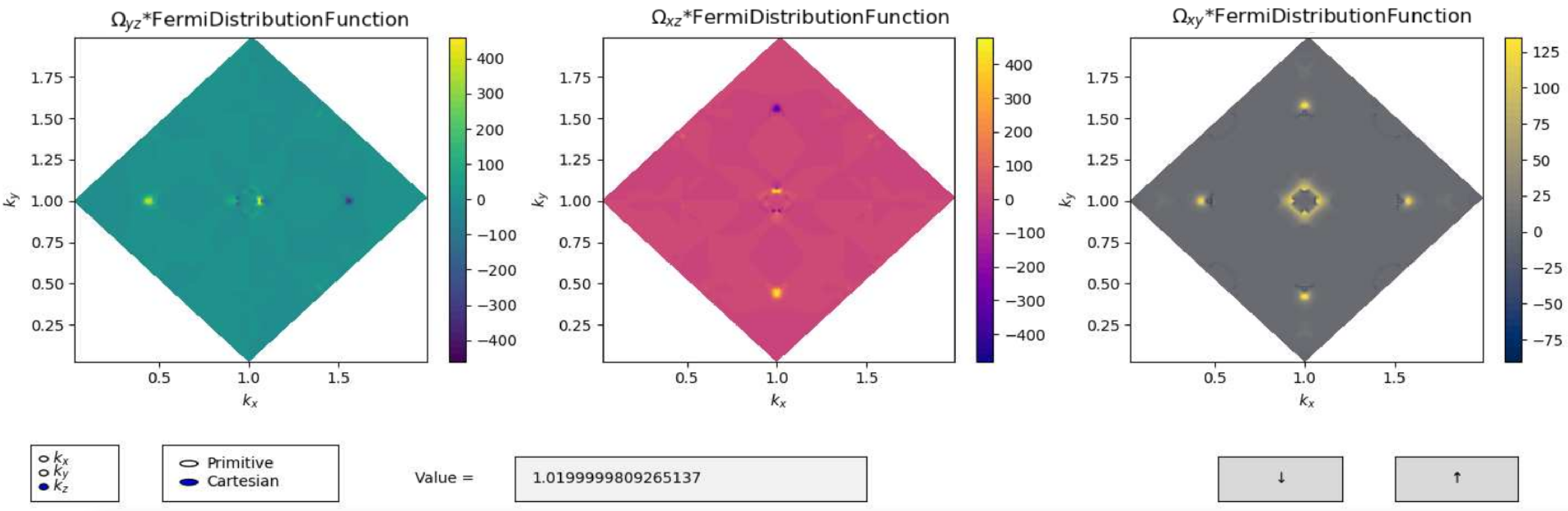}
    \caption{Interactive window produced by \textit{berryTrans\_plot.py} module for Fe during AHC computation at $T$ = 0 $K$ with the $\omega$ set to Fermi level. The three panels display the $\boldsymbol\Omega$ contributions- $\Omega_{yz}(\textbf{\textit{k}})*FermiDistributionFunction$, $\Omega_{xz}(\textbf{\textit{k}})*FermiDistributionFunction$ \& $\Omega_{xy}(\textbf{\textit{k}})*FermiDistributionFunction$, respectively, evaluated on a constant-$k_z$ plane at $k_z$= 1.0199999809265137 in the Cartesian coordinate system.}
    \label{berryPlotFe}
\end{figure*}

\section{The \textit{berryTrans\_plot.py} module}
It is seen from Eq. \ref{eqAHE} that the term $\sum_n f_n(\textbf{\textit{k}}) \Omega_{\mu \nu}^n(\textbf{\textit{k}})$ plays a vital role in deciding the extent to which a \textbf{\textit{k}}-point will contribute to total AHC (at a given value of $T$ and $\omega$) of any material under study. Similarly, Eq. \ref{eqANE} suggests that the term $\Omega_{\mu \nu}^n(\textbf{\textit{k}}) \left[ (E_n - E_F) f_n(\textbf{\textit{k}}) + k_B T \ln \left(1 + \exp\left( \frac{E_n - E_F}{-k_B T} \right) \right) \right]$ decides the extent to which a \textbf{\textit{k}}-point will contribute to total ANC (at a given value of $T$ and $\omega$) of any material which is explored.  In this regard, it is important to mention here that this contribution may get canceled out by the contribution from a term with equal magnitude but opposite sign at some other \textbf{\textit{k}}-point of the BZ. Thus, study of variation of these terms across the BZ may give some deeper insight regarding the origin of AHC/ANC in any given material. Such analysis is specially important in studying WSMs where AHC/ANC is claimed to get enhanced due to the states close to Weyl nodes\cite{shen2022intrinsically,zhou2020enhanced}. For such analysis, \textit{C-BerryTrans} is equipped with a python module named \textit{berryTrans\_plot.py}. This module must be used as post-processing tool after the calculation of $\boldsymbol\Omega$. Also, for just the visual analysis of the variation of $\boldsymbol\Omega$, the computation can be done on smaller size of \textbf{\textit{k}}-mesh (50$\times$50$\times$50 or 100$\times$100$\times$100) as compared to AHC/ANC computation. Various functions of this module and steps to use it is discussed further.

 Based on the value assigned to the \textit{property} variable, the module calculates $\Omega_{\mu \nu}(\textbf{\textit{k}})*FermiDistributionFunction$=$\sum_n f_n(\textbf{\textit{k}}) \Omega_{\mu \nu}^n(\textbf{\textit{k}})$ or $\Omega_{\mu \nu}(\textbf{\textit{k}})*EntropyDistributionFunction$=\\$\sum_n \Omega_{\mu \nu}^n(\textbf{\textit{k}}) \left[ (E_n - E_F) f_n(\textbf{\textit{k}}) + k_B T \ln \left(1 + \exp\left( \frac{E_n - E_F}{-k_B T} \right) \right) \right]$. Here, $\mu$, $\nu$=$x$, $y$ or $z$. The $T$ and $\omega$ value corresponding to which the calculation is to be done must be assigned to variables \textit{berryTrans\_plot\_at\_T} and \textit{chemical\_potential}, respectively. After this, copy the \textit{berryTrans\_plot.py} module into the working directory. The module will be available in the installation directory along with other C++ modules. Having this done, one must run the command- \textit{python3 berryTrans\_plot.py} to perform the calculations. After completing the computations, it generates an interactive window displaying  $\Omega_{yz}(\textbf{\textit{k}})*FermiDistributionFunction$, $\Omega_{xz}(\textbf{\textit{k}})*FermiDistributionFunction$ \& $\Omega_{xy}(\textbf{\textit{k}})*FermiDistributionFunction$ in $k_x$=$C$ or $k_y$=$C$ or $k_z$=$C$ plane, when \textit{property}=1. Similarly, if \textit{property}=2, it displays $\Omega_{yz}(\textbf{\textit{k}})*EntropyDistributionFunction$, $\Omega_{xz}(\textbf{\textit{k}})*EntropyDistributionFunction$ \& $\Omega_{xy}(\textbf{\textit{k}})*EntropyDistributionFunction$. Here, the parameter $C$ denotes a user-defined constant specifying the location of the two-dimensional momentum-space slice, such that $\boldsymbol\Omega$-driven transport quantities are visualized on planes of the form $k_x = C$, $k_y = C$, or $k_z = C$. At any given time, the code visualizes data on only one such plane; however, the user can switch to a different plane interactively within the plotting window using the available options. The screenshot of the window obtained on performing the AHC calculations on Fe at 0 \textit{K} with the $\omega$ value is set at Fermi level, is shown in Fig. \ref{berryPlotFe}. The three panels shown in figure correspond to the contributions from three Cartesian components of the $\boldsymbol\Omega$ evaluated on the same momentum-space plane. As can be seen in the figure, the window provides various interactive options for further analysis of $\Omega_{\mu \nu}(\textbf{\textit{k}})$. For instance, it provides the option to switch between $k_x$=$C$, $k_y$=$C$, and $k_z$=$C$ planes. In addition to this, it also provides an option to change the value of $C$ using the arrow key as shown in the figure. This flexibility enables the analysis of physically relevant planes, such as BZ boundary planes in nonsymmorphic materials or $k_z = 0$ and $k_z = \pi$ planes in hexagonal systems. The value of $C$ at which $\Omega_{\mu \nu}(\textbf{\textit{k}})$ term is shown at any instance is displayed in the text box with label \textit{Value}. Apart from this, one often needs to switch from primitive coordinate system to cartesian coordinate system or vice-versa. For instance, if one has the coordinates of Weyl nodes in either primitive coordinate system or cartesian coordinate system and need to verify how $\boldsymbol\Omega$ behaves around the Weyl node. Keeping these aspects in mind, the window is also provided with an option to switch from primitive to cartesian coordinate system or vice-versa. One must note that this conversion is only implemented for the structures mentioned in table \ref{tab72}. The features provided by \textit{berryTrans\_plot.py} module is expected to be very useful for $\boldsymbol\Omega$ \& AHC/ANC analysis in any material and to have a better understanding of the origin of the properties in the material. Thus, the module greatly enhance the capability of \textit{C-BerryTrans} package in exploring topological materials.

\section{Conclusion}

In the present work, a C++ based code named \textit{C-BerryTrans} has been designed to calculate intrinsic (i.e., $\boldsymbol\Omega$-driven) AHC or ANC of any material using the \textit{first-principles} approach. The code employs Kubo-like formula to calculate the $\boldsymbol\Omega$ by using the eigenvalues and momentum-matrix elements obtained from WIEN2k calculations. Parallel computing of $\boldsymbol\Omega$ \& the anomalous transport coefficients (AHC \& ANC) along with the efficient handling of band-resolved $\boldsymbol\Omega$ in binary format is found to greatly enhance the efficiency and usability of the code. For benchmarking of the code, it has been tested on some well-studied ferromagnetic materials exhibiting AHC or ANC. It has been tested on Fe, Fe$_3$Ge \& Co$_2$FeAl for AHC while on Pd, Fe$_3$Al \& Co$_2$FeAl for ANC. The results obtained, calculated from \textit{C-BerryTrans} code, are found to be in good match with those reported in literature. For instance, for the case of Fe, the magnitude of $\sigma_{xy}^{AHC}$ when chemical potential ($\omega$) is set at the Fermi energy is obtained to be $\sim$775 $S/cm$ at 0 $K$, which is found to be in close agreement to previously reported theoretical data (751 $S/cm$ at 0 $K$\cite{yao2004first}). Moving further, in case of Fe$_3$Ge, the calculated value of $\sigma_{xy}^{AHC}$ is found to be 311 $S/cm$ at the room temperature. This result is also found to be in a good match with the previously reported wannierization-based calculated result (227 $S/cm$) at 300 $K$\cite{li2023anomalous}. Nextly, for Co$_2$FeAl, the magnitude of computed value of $\sigma_{xy}^{AHC}$ at 2 $K$ when the value of $\omega$ is set at the Fermi level is found to be $\sim$56 $S/cm$. This is also in good agreement with the previously reported theoretical result (42 $S/cm$\cite{shukla2022atomic}, 39 $S/cm$\cite{huang2015anomalous}) and the experimental data (155 $S/cm$\cite{shukla2022atomic}) at 2 $K$. Furthermore, the room temperature magnitude of $\alpha_{xy}^{ANC}$ for Pd is obtained to be 0.97 $AK^{-1}m^{-1}$ which is in good match with the previously reported data (0.72 $AK^{-1}m^{-1}$\cite{guo2014anomalous}). In case of Fe$_3$Al, the maximum magnitude of $\alpha_{xy}^{ANC}$ for $T\leq$500 $K$ is computed as 2.83 $AK^{-1}m^{-1}$ which is also in good agreement with previously reported works (3.0 $AK^{-1}m^{-1}$\cite{koepernik2023symmetry},2.7 $AK^{-1}m^{-1}$\cite{sakai2020iron}). Lastly, for Co$_2$FeAl, the value of $\alpha_{xy}^{ANC}$ is obtained to be $\sim$0.10 $AK^{-1}m^{-1}$ at 300 $K$ which is in fairly good match with the data available in literature (0.06 $AK^{-1}m^{-1}$ at 300 $K$\cite{noky2020giant}). These findings confirm the precision, computational efficiency, and robustness of the \textit{C-BerryTrans} code in evaluating the AHC and/or ANC for a wide range of materials. Furthermore, the capability to compute $\boldsymbol\Omega$ along user-defined \textbf{\textit{k}}-point paths with appropriate Fermi–Dirac or entropy weighting provides direct insight into the interplay between band structure features and $\boldsymbol\Omega$-driven transport responses. Additionally, the \textit{berryTrans\_plot.py} module provided with the package further enhance the capability of the code in efficient exploration of the topological materials.

\bibliographystyle{apsrev4-2}
\bibliography{ref} 

\end{document}